\documentclass[prd,a4paper,nofootinbib,twocolumn,final,floatfix]{revtex4-2}
\usepackage{amsmath}
\usepackage{amssymb}
\usepackage{graphicx}
\usepackage[usenames,dvips]{color}
\usepackage[english]{babel}
\usepackage[latin1]{inputenc}
\usepackage{amsfonts}
\usepackage{appendix}
\usepackage{epsfig}
\usepackage{graphics,rotating}
\usepackage{dcolumn}
\usepackage{bm}
\usepackage{epstopdf}
\usepackage{color}
\usepackage[usenames,dvipsnames,svgnames]{xcolor}
\usepackage[colorlinks=true,
            linkcolor=blue,
            urlcolor=black,
            citecolor=magenta]{hyperref}
\usepackage{hyperref}
\usepackage[T1]{fontenc}
\usepackage{multirow}
\usepackage{float}
\usepackage{subfigure}
\usepackage{enumitem}
\usepackage{thmtools}

\begin{document}

\title{Spin precession of binary neutron stars with magnetic dipole moments}
\author{Bal\'azs Mik\'oczi}

\affiliation{Research Institute for Particle and Nuclear Physics,
Wigner RCP H-1525
Budapest 114, P.O. Box 49, Hungary\\
{\small E-mail: mikoczi.balazs@wigner.hu\quad }}

\begin{abstract}
Spin precession equations including the spin-orbit (SO), spin-spin (SS),
quadrupole-monopole (QM) and magnetic dipole-magnetic dipole (DD)
leading-order interactions are derived for compact binary systems in order
to investigate the DD contribution in the orbit-averaged spin precession
equations for binary neutron star systems neglecting the gravitational
radiation-reaction effect. It is known that the magnitudes of spins are not
conserved quantities due to the DD interaction. We give a simple analytical
description for the pure DD interaction making the magnitudes of spins
almost constant by neglecting the SO, SS and QM contributions. We also
demonstrate the evolutions of the relative angles of spins and magnetic
dipoles with the help of numeric simulations including all contributions
(SO, SS, QM and DD) and introduce a dimensionless magnetic dipole parameter
to characterize the strength of magnetic fields for some realistic neutron
star binaries. We find that for realistic configurations the strong magnetic
fields of neutron stars can modify the spin dynamics over long periods of
time.
\end{abstract}

\maketitle

\section{Introduction}

Nowadays, spinning astrophysical objects have come to the forefront in
particular through the detection of gravitational waves (GWs). The most
important targets of GW detectors are binaries of neutron stars (NS) and/or
black holes (BH) \cite{LIGO1,LIGO2,LIGO3,LIGO4,LIGO5NS,LIGO6,LIGO7,LIGO8}.
Numerous independent astronomical observations have also confirmed that
spins of objects are not negligible, e.g., spin magnitude\footnote{%
Definition of the dimensionless spin parameter is $\chi =cS/Gm^{2}$%
, where $G$ is the gravitational constant, $c$ is the speed of light, $S$ is
the spin magnitude and $m$ is the mass.} can be as high as $0.9$ in case of
the recently discovered supermassive M87 BH \cite{M87}. Several studies have
shown the spin of a BH to be commonly close to $1$, which is the case of an
extremal Kerr BH \cite{Gou2011}. The maximum spin value of a NS is $0.7$ for
a wide class of realistic equations of state \cite{WaiLo2011}. But the
realistic spin of a NS is rather small, for instance, the fastest known
pulsar in a binary NS system, J0737-3039A has a value of $0.03$. In the
process of GW searches there is a non-trivial loss in signal-to-noise ratio
if the maximum spin value is restricted to be less than $0.05$ \cite%
{Brown2012}. Numerical studies have shown that in the case of a minor
residual spin of $10^{-4}$ the evolution of spinning NS binaries indicates
that their spin precession can be well described by post-Newtonian
approximation \cite{Tacik2015}.

Consideration of spins $\mathbf{S_{i}}$ of bodies in physical
systems leads to spin precession equations (SPEs) which are
important for the investigation of classical and/or quantum systems.
The first analytical solution of SPEs averaged over one orbital
period (ASPEs) has been given for compact binaries containing the
leading-order spin-orbit (SO) interaction and the leading-order
radiation-reaction contribution only \cite{ACST}. They have
presented the transitional and the simple precession movements of
the orbital plane due to spin dynamics for two equivalent cases,
which are the equal mass and the single spin configurations. Further
interpretations and applications of this approach were given for
spin-flip and spin flip-flop effects \cite{flipflop}. ASPEs with
spin-spin (SS) contributions taking into account gravitational
radiation were also analytically solved for the case of equal mass
and equal spin magnitude \cite{Apostolatos96}.

It is important to note that in case of BHs \cite{Racine} ASPEs can be
integrated by adding the quadrupole-monopole (QM) terms, where the
dimensionless quadrupole parameters\footnote{%
Quadrupole parameters $a_{i}$ strongly depend on the assumed equation of
state and are $a_{i}=1$ for BHs, $a_{i}=2..14$ for NSs and $a_{i}=10..150$
for boson stars.} $a_{i}$ are equal to $1$ \cite{Poisson}. In this case
there is a new conserved quantity $\xi $\footnote{%
The new conserved quantity for SO, SS and QM contributions is $\xi =\mathbf{%
L_{N}}\cdot \mathcal{S}\mathbf{/}|\mathbf{L_{N}}|^{2}$, where $\mathbf{L_{N}=%
}\mu \mathbf{r\times v}$ is the Newtonian orbital momentum, $\mu
=m_{1}m_{2}/(m_{1}+m_{2})$ is the reduced mass, $\mathbf{r}$ is the relative
position vector and $\mathbf{v}$ is the relative velocity. The quantity $%
\mathcal{S}=\left( 1+m_{2}/m_{1}\right) \mathbf{S_{1}}+\left(
1+m_{1}/m_{2}\right) \mathbf{S_{2}}$ consists of the masses $m_{i}$ and the
individual spin vectors $\mathbf{S_{i}}$.} on which many further spin
precession analytical solutions are built \cite%
{Kesden2015,Gerosa2015,Gerosa2019,Chatziioannou2017,Khan2019}. It is also
important to mention that if the quadrupole parameters are not equal to $1$ (%
$a_{i}\neq 1$), i.e., in the case of NSs, this $\xi $ scalar is not a
conserved quantity.

Solution of the instantaneous SPEs with SO, SS and QM terms for
arbitrary quadrupole parameters was given by perturbative methods
\cite{majarmikoczi}, where the motion of the polar and azimuthal
angles can be completely separated. It is interesting to note that
the SPEs do not depend on the spin supplementary conditions (SSC),
while the radial and angular motion and waveforms are SSC-dependent
in case of a compact binary with leading order spin-orbit
interaction \cite{Mikoczi2017}. Equilibrium solutions of the SPEs
can be found in \cite{Schnittman}. Secular SPEs for NSs, gravastars
and boson stars, where the quadrupole parameters are also not equal
to 1, have been studied in \cite{Gergely1} and their linear
stability was analyzed in \cite{Gergely2}. Recently, higher-order
SPEs were also examined analyzing numerically generated
gravitational waveforms and hybrid models \cite{Akcay2021}.

The observed radio pulsar NSs have strong magnetic fields typically of $%
10^{12}$ G, while magnetars have fields of $10^{12}-10^{15}$ G. The recently
discovered youngest and fastest magnetar Swift J1818.0-1607 has a magnetic
field of $7\times 10^{14}$ G and a spin period of $1.36$ s \cite{Swift}.
Therefore it is important to examine whether the strength of this magnetic
field or dipole field plays a role in spin dynamics. The Lagrangian
formalism of the magnetic dipole-magnetic dipole (DD) interaction, the
energy and angular momentum losses under gravitational radiation in case of
circular and eccentric orbits are given in \cite{IT,DD,MVG}. The generalized
Kepler equation containing the DD contribution has been calculated in \cite%
{KMG}, but the DD interaction generates effects of the same magnitude as the
second post-Newtonian correction, in pure relativistic terms, in magnetic
fields up to $10^{16}$ G. Recently, some authors pointed out that the
magnitude of the magnetic dipole moment can also be larger by order of three
magnitudes for white dwarfs than for NSs \cite{Bourgoin2021}. They also
estimated the precession rates of the magnetic moments for azimuthal angles
using a model where the magnetic dipoles are aligned with the corresponding
spin vectors and showed that the DD effect in GW could be important for
future data processing of the Laser Interferometer Space Antenna (LISA)
mission. It is worth to note that the full description of spinning binary
NSs with considerable magnetospheres is given by magnetohydrodynamic (MHD)
equations. Some MHD simulations demonstrated the conversion of a binary
system's kinetic energy into electromagnetic radiation through unipolar
induction and accelerating magnetic dipole effects \cite{CarrascoShibata2020}%
.

In this paper, we study the influence of the magnetic field on the
conservative part of ASPEs due to DD contribution through the inspiral
phase, while we do not take into account the gravitational radiation effect.
In this case, the magnitudes of the spin vectors are not constants.
Therefore $\xi $ will not be a conserved quantity, which means that an
analytical approach similar to \cite{Racine} will not work in this case. The
DD contribution is usually rather small, thus it is important to investigate
the long-term evolution of a binary NS system. Initially, we neglect the
standard effects arising from the SO, SS and QM contributions and instead
focus only on the DD interaction, which model we will call the \textit{pure
DD} case further on. We solve this model for a simple case, when we only
consider the linear-order magnetic dipole terms. The validity of this model
is examined by numerical developments. Next, we numerically examine the
solution of the total ASPEs with all contributing terms (SO, SS, QM, and DD)
included. We demonstrate the time evolution of relative spin angles through
the inspiral phase for several binary NS systems with realistic spin and
magnetic dipole ratios (Table \Ref{table1}). Furthermore, we present some
cases for which it is important to take into account the magnetic dipoles of
binary NSs.

The paper is organized as follows. In Section 2 we introduce the SPEs
containing the SO, SS, QM and DD contributions. In Section 3 we study the
pure DD model in SPEs neglecting the SO, SS and QM interactions. We compare
the perturbative solutions of the pure DD model with the exact numeric
solutions and present the limits of applicability for the perturbative
method in this case. In Section 4 the discussion of the SPEs, containing all
contributions, based on various numerical examples is presented and Section
5 contains the conclusion.

Throughout the paper we use the geometric unit system in which $G=c=\mu
_{0}=1$, where $G$ is the gravitational constant, $c$ is the speed of light
and $\mu _{0}$ is the vacuum permeability ($\mu _{0}=4\pi \times 10^{-7}$
Tm/A in SI units), but we kept the SI units in Table \Ref{table1}. We will
not use the Einstein convention thus there is no summation for repeated
indices. The overhat symbol above a vector represents the unit vector
notation, e.g., for any vector $\mathbf{A}$ then $\mathbf{\hat{A}=A/}A$ is
the unit vector, where $A=\left\vert \mathbf{A}\right\vert $ is the
magnitude of $\mathbf{A}$.

\section{Spin precession equations}

\begin{figure}[tbh]
\begin{center}
\includegraphics[width=0.4\textwidth]{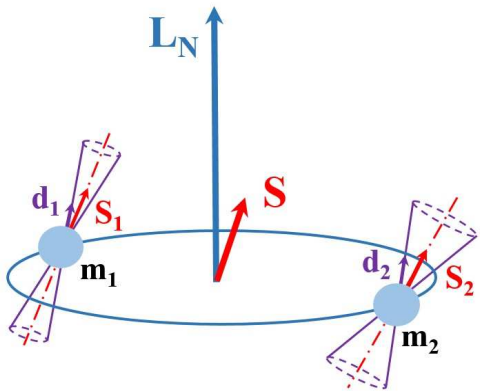}
\end{center}
\caption{Geometry of a binary NS. The orbital plane is determined by the
Newtonian orbital momentum $\mathbf{L_{N}}$. The characteristics of the
individual NSs are their masses $m_{i}$, spins $\mathbf{S_{i}}$ and magnetic
dipole moments $\mathbf{d_{i}}$.}
\label{geometry}
\end{figure}

We consider a binary NS system where the characteristics of the individual
objects are their masses $m_{i}$, mass moments of inertia $\mathcal{I}_{i}$,
quadrupole parameters $a_{i}$, spin vectors $\mathbf{S_{i}} $ and magnetic
dipole moments $\mathbf{d}_{\mathbf{i}}$ (see Fig. \ref{geometry}). The
radial motion and gravitational radiation effects of the DD interaction can
be found in \cite{IT,DD,MVG}. We review the conservative part of the
orbit-averaged SPE system without radiation-reaction contributions, which
are given by \cite{ACST,DD,Racine}
\begin{eqnarray}
\mathbf{\dot{S}_{1}} &=&\frac{1}{2r^{3}}\Bigl\{\left( 4+3\nu _{1}\right)
\mathbf{L_{N}\times \mathbf{\mathbf{S_{1}}}+S_{2}\times \mathbf{S_{1}}}
\notag \\
&&-3\left[ \mathbf{\hat{L}_{N}\cdot S_{2}}+a_{1}\nu _{1}(\mathbf{\hat{L}%
_{N}\cdot S_{1}})\right] (\mathbf{\hat{L}_{N}\times \mathbf{S_{1})}}  \notag
\\
&&\ \mathbf{+d}_{2}\times \mathbf{d_{1}+}3(\mathbf{d}_{2}\mathbf{\cdot \hat{L%
}}_{\mathbf{N}})(\mathbf{\hat{L}}_{\mathbf{N}}\times \mathbf{d_{1})}\Bigr\}\
,  \label{S1dot} \\
\mathbf{\dot{S}_{2}} &=&\frac{1}{2r^{3}}\Bigl\{\left( 4+3\nu _{2}\right)
\mathbf{L_{N}\times \mathbf{S_{2}}+S_{1}\times \mathbf{S_{2}}}  \notag \\
&&\mathbf{\mathbf{-}}3\left[ \mathbf{\hat{L}_{N}\cdot S_{1}}+a_{2}\nu _{2}(%
\mathbf{\hat{L}_{N}\cdot S_{2}})\right] (\mathbf{\hat{L}_{N}\times \mathbf{%
S_{2})}}  \notag \\
&&\mathbf{+d}_{1}\times \mathbf{d_{2}+}3(\mathbf{d}_{1}\mathbf{\cdot \hat{L}}%
_{\mathbf{N}})(\mathbf{\hat{L}}_{\mathbf{N}}\times \mathbf{d_{2})}\Bigr\}\ ,
\label{S2dot} \\
\mathbf{\dot{L}_{N}} &=&\frac{1}{2r^{3}}\Bigl\{(4\mathbf{S}+3\bm{\sigma }%
)\times \mathbf{L_{N}}  \notag \\
&&\mathbf{-}3\left[ (\mathbf{\hat{L}}_{\mathbf{N}}\mathbf{\cdot S}_{\mathbf{2%
}})\mathbf{S}_{\mathbf{1}}+(\mathbf{\hat{L}}_{\mathbf{N}}\mathbf{\cdot S}_{%
\mathbf{1}})\mathbf{S}_{\mathbf{2}}\right] \times \mathbf{\hat{L}}_{\mathbf{N%
}}  \notag \\
&&+{3}\sum_{i=1}^{2}a_{i}\nu _{i}(\mathbf{\hat{L}}_{\mathbf{N}}\mathbf{\cdot
S_{i}})(\mathbf{\hat{L}}_{\mathbf{N}}\mathbf{\times \mathbf{S_{i})}}  \notag
\\
&&\mathbf{\mathbf{+}}3\sum_{i\neq j}^{2}\mathbf{(d_{i}\cdot \hat{L}%
_{N})(d_{j}\times \hat{L}_{N}})\Bigr\}\ ,  \label{LNdot} \\
\mathbf{\dot{d}_{i}} &=&\frac{1}{\mathcal{I}_{i}}\mathbf{S}_{\mathbf{i}%
}\times \mathbf{d_{i}}\ ,  \label{ddot}
\end{eqnarray}%
where the overdot represents differentiation with respect to time, $%
r=a(1-e^{2})^{1/2}$ is the orbit-averaged diameter, $a$ is the semimajor
axis and $e$ is the eccentricity (for further details see \cite{BOC}). In
addition, $r$ can also be considered as the radius of the circular orbit
(e.g., in \cite{ACST}). The Newtonian orbital momentum is $\mathbf{L_{N}=}%
\mu \mathbf{r\times v}$, where $\mu =m_{1}m_{2}/(m_{1}+m_{2})$ is the
reduced mass, $\mathbf{r}$ is the relative position vector and $\mathbf{v}$
is the relative velocity. The unit vector of $\mathbf{L_{N}}$ is $\mathbf{%
\hat{L}_{N}=L_{N}/}L_{N}$\textbf{, }where $L_{N}=\left\vert \mathbf{L_{N}}%
\right\vert $ is the magnitude of $\mathbf{L_{N}}$. Furthermore, the
shorthand notations are $\nu _{1}=m_{2}/m_{1}$,\ $\nu _{2}=m_{1}/m_{2}$, $%
\mathbf{S=S}_{\mathbf{1}}+\mathbf{S}_{\mathbf{2}}$ and $\bm{\sigma =}\nu _{1}%
\mathbf{S}_{\mathbf{1}}+\nu _{2}\mathbf{S}_{\mathbf{2}}$. There is no
summation for repeated indices $i$ in Eq. (\ref{ddot}).

The linear terms of spins in equation system Eqs. (\ref{S1dot}-\ref{LNdot})
are the SO contributions. The QM terms are proportional to the quadrupole
parameter $a_{i}$ and are quadratic in spin, while the other
quadratic-in-spin terms are the SS interactions. Obviously, the DD terms in
SPEs are proportional to the product of dipole moments and the evolution
equations of dipole moments are represented by Eq. (\ref{ddot}).

The conserved quantities in general case are the total angular momentum
vector $\mathbf{J=L_{N}+S}$ (here we neglect the spin-orbit angular momentum
term $\mathbf{L}_{\mathbf{SO}}$ and higher order post-Newtonian
contributions), the magnitude of the Newtonian angular momentum $L_{N}$ and
the magnitudes of the magnetic dipole moments $d_{i}$. The magnitudes of
spin vectors $S_{i\text{ }}=\left\vert \mathbf{S}_{\mathbf{i}}\right\vert $
are not conserved quantities and from Eqs. (\ref{S1dot},\ref{S2dot})%
\begin{eqnarray}
\dot{S}_{1} &=&\frac{d_{1}d_{2}}{2r^{3}}(\mathbf{\hat{d}_{1}}\times \mathbf{%
\hat{S}_{1})\cdot }\left[ \mathbf{\hat{d}_{2}+}3(\mathbf{\hat{d}_{2}\cdot
\hat{L}}_{\mathbf{N}})\mathbf{\hat{L}}_{\mathbf{N}}\right] ,  \label{S1mag}
\\
\dot{S}_{2} &=&\frac{d_{1}d_{2}}{2r^{3}}(\mathbf{\hat{d}_{2}}\times \mathbf{%
\hat{S}_{2})\cdot }\left[ \mathbf{\hat{d}_{1}+}3(\mathbf{\hat{d}_{1}\cdot
\hat{L}}_{\mathbf{N}})\mathbf{\hat{L}}_{\mathbf{N}}\right] ,  \label{S2mag}
\end{eqnarray}%
where we used the unit vector notations of $\mathbf{\hat{S}_{i}=S}_{\mathbf{i%
}}/S_{i}$ and $\mathbf{\hat{d}_{i}=}$ $\mathbf{d_{i}/}d_{i}$ with magnitude
of the dipole moments $d_{i}=\left\vert \mathbf{d_{i}}\right\vert $. It can
be seen that the time evolutions of magnitudes of individual spins are
rather small $\dot{S}_{i}\approx \mathcal{O}(d_{1}d_{2}/r^{3})$, but they
can have a significant effect in the long-term evolution of NSs.

The $\xi =\mathbf{L_{N}}\cdot \bm{(S+\sigma)}/|\mathbf{L_{N}}|^{2}$
quantity, which is conserved for the SO, SS and QM contributions, is not a
conserved quantity due to the DD contribution, because in this case%
\begin{eqnarray}
\dot{\xi} &\propto &{3}\nu _{1}\left( a_{1}\mathbf{-}1\right) (\mathbf{\hat{L%
}}_{\mathbf{N}}\mathbf{\cdot S}_{\mathbf{1}})(\mathbf{\mathbf{\mathbf{S_{1}}}%
}\times \mathbf{S}_{\mathbf{2}})\mathbf{\cdot \mathbf{\hat{L}}_{\mathbf{N}}}
\notag \\
&&\mathbf{+}3\nu _{2}\left( a_{2}\mathbf{-}1\right) (\mathbf{\hat{L}}_{%
\mathbf{N}}\mathbf{\cdot S}_{\mathbf{2}})(\mathbf{\mathbf{\mathbf{S_{2}}}}%
\times \mathbf{S}_{\mathbf{1}})\mathbf{\cdot \hat{L}}_{\mathbf{N}}  \notag \\
&&\mathbf{\mathbf{-}}6\mathbf{(d_{1}\cdot \hat{L}_{N})(\hat{L}_{N}\times
d_{2}})\mathbf{\cdot \mathbf{\bm{(S+\sigma )}}}  \notag \\
&&\mathbf{\mathbf{-}}6\mathbf{(d_{2}\cdot \hat{L}_{N})(\hat{L}_{N}\times
d_{1}})\mathbf{\cdot \mathbf{\bm{(S+\sigma )}}}  \notag \\
&&\mathbf{-}2\nu _{1}\left( \mathbf{d}_{2}\times \mathbf{d_{1}}\right)
\mathbf{\cdot L_{N}\mathbf{-}}2\nu _{2}\mathbf{\left( \mathbf{d}_{1}\times
\mathbf{d_{2}}\right) \cdot L_{N},}
\end{eqnarray}%
which will only be zero if
\begin{eqnarray}
a_{i} &=&1\text{ (for BHs) \& }m_{1}\!=\!m_{2}\text{ \& }\mathcal{O}\mathbf{(%
}Sd^{2})\mathbf{\approx }0,  \label{case1} \\
a_{i} &\neq &1\text{ (for NSs) \& }m_{1}\!=\!m_{2}\text{ \& }\mathcal{O}%
\mathbf{(}S^{3},Sd^{2})\mathbf{\approx }0,  \label{case2} \\
a_{i} &\neq &1\text{ (for NSs) \& }m_{1}\!\neq \!m_{2}\text{ \& }\mathcal{O}%
\mathbf{(}S^{3},Sd^{2},d^{2})\mathbf{\approx }0.  \label{case3}
\end{eqnarray}%
It is worth to note that the last condition in (\ref{case1}) can also be
satisfied by adding an other quantity $\xi _{S^{2}}=\mathbf{S_{1}\cdot S_{2}}%
+\mathbf{S_{1}^{2}/}2\mathbf{+S_{2}^{2}/}2$ to $\xi $ and then the modified
scalar $\xi +\xi _{S^{2}}$ will be a conserved quantity for equal-mass BHs ($%
a_{i}=1$ and $m_{1}=m_{2}$). The system of equations in Eqs. (\ref{S1dot}-%
\ref{LNdot}) cannot be solved by the analytical method used in \cite{Racine}
even if we neglect the changes in the $\xi $ quantity ($\dot{\xi}\mathbf{%
\approx }0$), since magnetic dipole vectors result in complicated coupling
with spin vectors. This system of equations in Eqs. (\ref{S1dot}-\ref{LNdot}%
) has 9 degrees of freedom and integration of these equations is rather
complicated. Therefore we neglect the SO, SS and QM terms, which we shall
discuss in the next subsection.

\begin{widetext}

\begin{table}[tbp]
\begin{tabular}{ccccccc}
\hline\hline Physical parameters & J0737-3039A & J0737-3039B &
B1913+16 & J1930-1852 & J1906+0746 & J1755-2550 \\ \hline
\multicolumn{1}{l}{Masses, $m_{1}$, $m_{2}$ [$M_{\odot }$]} &
1.338,1.249 & 1.249,1.338 & 1.438,1.390 & <1.25,>1.30 & 1.291,1.322
& -,>0.40
\\
\multicolumn{1}{l}{Spin frequency, $\Omega $ (Hz)} & 44.054 & 0.361
& 16.949
& 5.405 & 6.944 & 3.175 \\
\multicolumn{1}{l}{Surface magnetic field strength, $B$ [$\times
10^{5}$ T]}
& 6.4 & 1590 & 22.8 & 58.5 & 1730 & 886 \\
\multicolumn{1}{l}{Mass moment of inertia, $\mathcal{I}$ [$\times
10^{38}$ kgm$^{2}
$]} & 1.064 & 0.994 & 1.144 & 0.994 & 1.027 & 0.318 \\
\multicolumn{1}{l}{Magnitude of spin, $S$ [$\times 10^{39}$
kgm$^{2}$/s]} & 4.689 &
0.036 & 1.939 & 0.538 & 0.713 & 0.101 \\
\multicolumn{1}{l}{Magnitude of magn. dip. mom., $d$ [$\times
10^{26}$ Am$^{2}$]}
& 0.021 & 5.30 & 0.076 & 0.195 & 5.766 & 2.953 \\
\multicolumn{1}{l}{Spin parameter, $\chi $ [$\times 10^{-4}$]} &
29.75 & 0.26
& 10.65 & 3.91 & 4.86 & 7.17 \\
\multicolumn{1}{l}{Magnetic dipole parameter, $\chi ^{d}$ [$\times
10^{-4}$]} &
0.006 & 0.160 & 0.002 & 0.01 & 10.16 & 0.862 \\
\multicolumn{1}{l}{Ratio, $\Delta =\chi ^{d}/\chi $ [$\times 10^{-2}
$]} & 0.02 & 60.8 & 0.02 & 0.1 & 3.3 & 12.0 \\
\hline\hline
\end{tabular}%
\caption{Measured and derived parameters for some binary NS
systems\textbf{.} The observed masses, spin frequencies and surface
magnetic fields are from the literature \protect\cite{DNS}. When
calculating the mass moment of
inertia $\mathcal{I}$, we considered the NS to be a regular sphere, so $%
\mathcal{I}=\frac{2}{5}mR^{2}$ with the typical radii of $R=10$ km.
Magnitudes of the spin and magnetic dipole moment are derived from the $S=%
\mathcal{I}\Omega $ and $d=VB/\protect\mu _{0}$ equations,
respectively,
where $V$ is the volume of the sphere and $\protect\mu _{0}=4\protect\pi %
\times 10^{-7}$ Tm/A is the vacuum permeability. The dimensionless
spin and
magnetic dipole parameters in SI units are $\protect\chi =cS/m^{2}G$ and $%
\protect\chi ^{d}=\protect\mu _{0}^{1/2}c^{2}d/m^{2}G^{3/2}$,
respectively.} \label{table1}
\end{table}

\end{widetext}

\section{Pure DD case}

\begin{figure}[tbh]
\begin{center}
\includegraphics[width=0.5\textwidth]{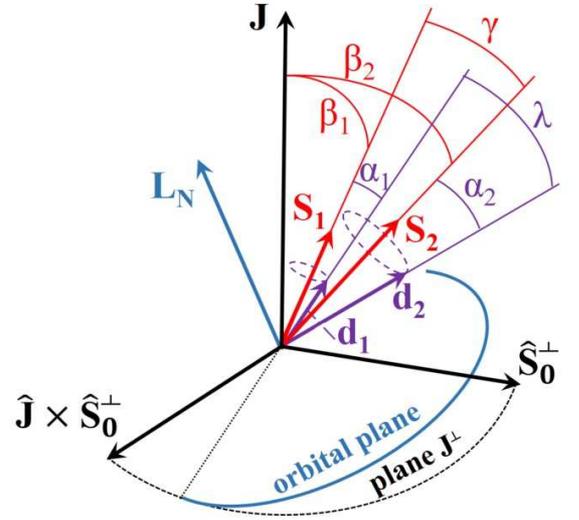}
\end{center}
\caption{Geometry of a binary NS with spins and magnetic dipole moments in
the coordinate system with the basis of $(\mathbf{\hat{S}_{0}^{\perp },\hat{J%
}}\times \mathbf{\hat{S}_{0}^{\perp },\hat{J}})$, where $\mathbf{%
S_{0}^{\perp }}$ is the projection of the initial total spin vector ($%
\mathbf{S=S_{1}}+\mathbf{S_{2}}$) onto the plane perpendicular to $\mathbf{J}
$, which plane is $\mathbf{J}^{\perp }$. The nodus is the dotted line which
is the intersection of the orbital plane and plane $\mathbf{J}^{\perp }$.
The definitions of the introduced relative angles $\protect\beta _{i}$, $%
\protect\alpha _{i}$, $\protect\gamma $ and $\protect\lambda $ of $\mathbf{J}
$, $\mathbf{S_{i}}$ and $\mathbf{d_{i}}$ are $\cos \protect\beta _{i}=%
\mathbf{\hat{J}\cdot \hat{S}_{i}}$, $\cos \protect\gamma =\mathbf{\hat{S}}_{%
\mathbf{2}}\mathbf{\cdot \hat{S}}_{\mathbf{2}},\cos \protect\alpha _{i}=%
\mathbf{\hat{S}_{i}\cdot \hat{d}_{i}}$ and $\cos \protect\lambda =\mathbf{%
\hat{d}_{1}\cdot \hat{d}_{2}}$. Additional definitions of polar angles,
which we use in the paper, not shown in the figure, are $\cos \protect\delta %
_{i}=\mathbf{\hat{J}\cdot \hat{d}_{i}}$, $\cos \protect\varrho _{i}=\mathbf{%
\hat{L}}_{\mathbf{N}}\mathbf{\cdot \hat{d}_{i}}$ and $\cos \protect\kappa %
_{i}=\mathbf{\hat{L}}_{\mathbf{N}}\mathbf{\cdot \hat{S}}_{\mathbf{i}}$.}
\label{spingeometry}
\end{figure}

First, we consider the DD interactions only in the SPEs in Eqs. (\ref{S1dot}-%
\ref{ddot}) \ by dropping the SO, SS and QM contributions, which is called
the \textit{pure DD} case, as
\begin{eqnarray}
\mathbf{\dot{S}_{1}} &=&\frac{1}{2r^{3}}\left[ \mathbf{d_{2}+}3(\mathbf{%
d_{2}\cdot \hat{L}}_{\mathbf{N}})\mathbf{\hat{L}}_{\mathbf{N}}\right] \times
\mathbf{d_{1}}\ ,  \label{DDpure1} \\
\mathbf{\dot{S}_{2}} &=&\frac{1}{2r^{3}}\left[ \mathbf{d_{1}+}3(\mathbf{%
d_{1}\cdot \hat{L}}_{\mathbf{N}})\mathbf{\hat{L}}_{\mathbf{N}}\right] \times
\mathbf{d_{2}}\ ,  \label{DDpure2} \\
\mathbf{\dot{L}_{N}} &=&\frac{3}{2r^{3}}\sum_{i\neq j}^{2}\mathbf{%
(d_{i}\cdot \hat{L}_{N})(d_{j}\times \hat{L}_{N}})\ ,  \label{DDpure3} \\
\mathbf{\dot{d}_{i}} &=&\frac{1}{\mathcal{I}_{i}}\mathbf{S}_{\mathbf{i}%
}\times \mathbf{d_{i}}\ .  \label{DDpure4}
\end{eqnarray}%
It can be seen from Eqs. (\ref{DDpure1},\ref{DDpure2},\ref{DDpure4}) that
the scalar products of $\mathbf{\mathbf{\mathbf{S_{i}\cdot \mathbf{d}_{i}}}}$
are conserved quantities. Therefore $\frac{d}{dt}(S_{i}d_{i}\cos \alpha
_{i})=0,$ where $\cos \alpha _{i}=\mathbf{\mathbf{\mathbf{\hat{S}_{i}\cdot
\hat{d}_{i}}}}$ (see Fig. \ref{spingeometry}), which are not conserved
quantities leading to the simple solutions of
\begin{equation}
\cos \alpha _{i}=\frac{C}{S_{i}},
\end{equation}%
where $C$ is an integration constant. It is worth to note that the polar
angles $\alpha _{i}$ are conserved quantities if we drop the evolutions of
the magnitudes of the spin vectors ($\dot{S}_{i}\approx 0$) in Eq. (\ref%
{S1mag}), in which case the magnetic dipoles would exhibit pure precessions
around the individual spins.

Thus, there are 8 conserved quantities ($\mathbf{J}$, $L_{N}\,$, $d_{i}$ and
$\mathbf{\mathbf{\mathbf{S_{i}\cdot \mathbf{d}_{i}}}}$) for the 15 ASPEs in
Eqs. (\ref{DDpure1}-\ref{DDpure4}). Therefore, the pure DD system has 7
degrees of freedom.

\subsubsection{Perturbative solution of pure DD case}

\begin{figure}[tbh]
\begin{center}
\includegraphics[width=0.45%
\textwidth]{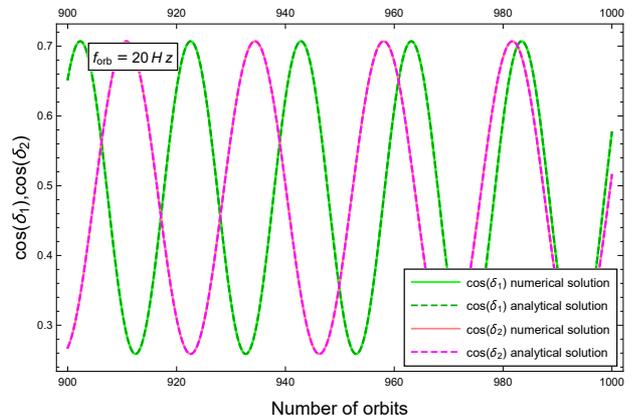}
\end{center}
\caption{Evolutions of relative angles $\cos \protect\delta _{i}$. Masses of
the binary NS are $1.4M_{\odot }$ and $1.2M_{\odot }$, while the orbital
frequency is $f_{orb}=20$ Hz. The dimensionless spin parameter is $\protect%
\chi =0.001$ ($\protect\chi _{1}=$ $\protect\chi _{2}\equiv \protect\chi $)
and the ratio parameter is $\Delta $ ($=\protect\chi ^{d}/\protect\chi =0.1$%
). Initial spin and dipole angles in $\mathcal{K}$ are chosen as $\protect%
\theta _{1}(0)=\protect\pi /4=\protect\theta _{2}(0)$, $\protect\varphi %
_{1}(0)=0$, $\protect\varphi _{2}(0)=\protect\pi /2$, $\protect\theta %
_{1}^{d}(0)=\protect\theta _{1}(0)+\protect\pi /12=\protect\theta %
_{2}^{d}(0) $, $\protect\varphi _{1}^{d}(0)=\protect\varphi _{1}(0)$ and $%
\protect\varphi _{2}^{d}(0)=\protect\varphi _{2}(0)$. The curves are plotted
for the last $100$ orbital periods ($t_{final}=1000T_{orb}$, where $%
T_{orb}=1/f_{orb} $).}
\label{evol0}
\end{figure}

\begin{figure}[tbh]
\begin{center}
\includegraphics[width=0.45\textwidth]{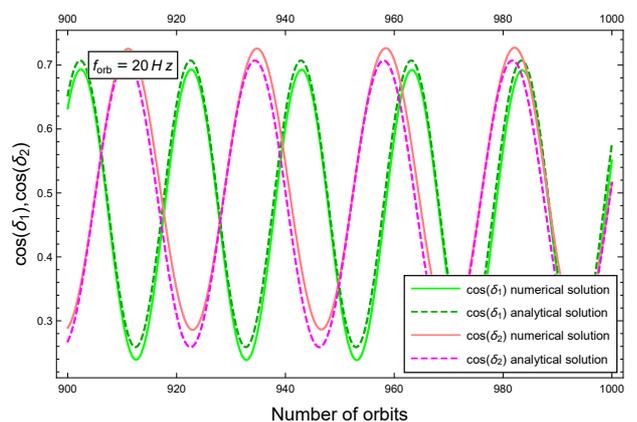}
\end{center}
\caption{Evolutions of relative angles $\cos \protect\delta _{i}$. The
parameters and initial values are the same as in Fig. \protect\ref{evol0}
with the exception of the ratio parameter which is $\Delta =5$. The
difference between the perturbative and numerical solutions of the DD case
is small with a maximum difference in amplitude of $2\%$.}
\label{evol02}
\end{figure}

In many cases, the ASPEs written in the system fixed to $\mathbf{J}$ led to
an analytical solution \cite{Apostolatos96,Racine,majarmikoczi}, since $%
\mathbf{J}$ is a conserved quantity when the radiation-reaction effect is
neglected. Therefore, in Eqs. (\ref{DDpure1},\ref{DDpure2},\ref{DDpure4}),
we consider the evolution equations for scalar products of $\mathbf{\hat{J}%
\cdot \hat{S}_{i}}$ and $\mathbf{\hat{J}\cdot \hat{d}_{i}}$ with total
angular momentum unit vector $\mathbf{\hat{J}=J/}J$ and the magnitude of the
total angular momentum $J=\left\vert \mathbf{J}\right\vert $ as
\begin{eqnarray}
\frac{d}{dt}(\mathbf{\hat{J}\cdot \hat{S}_{1})\!} &=&\!\frac{d_{1}d_{2}}{%
2r^{3}S_{1}}\Bigl\{\lbrack \mathbf{\hat{d}}_{\mathbf{2}}\mathbf{+}3(\mathbf{%
\hat{d}}_{\mathbf{2}}\mathbf{\cdot \hat{L}}_{\mathbf{N}})\mathbf{\hat{L}}_{%
\mathbf{N}}]\mathbf{\cdot }[(\mathbf{\hat{d}_{1}}\times \mathbf{\hat{J})}
\notag \\
&&\!+(\mathbf{\hat{d}_{1}}\times \mathbf{\hat{S}_{1})}(\mathbf{\hat{J}\cdot
\hat{S}_{1})}]\Bigr\}, \\
\frac{d}{dt}(\mathbf{\hat{J}\cdot \hat{S}_{2})\!} &=&\!\frac{d_{1}d_{2}}{%
2r^{3}S_{2}}\Bigl\{\lbrack \mathbf{\hat{d}}_{\mathbf{1}}\mathbf{+}3(\mathbf{%
\hat{d}}_{\mathbf{1}}\mathbf{\cdot \hat{L}}_{\mathbf{N}})\mathbf{\hat{L}}_{%
\mathbf{N}}]\mathbf{\cdot \lbrack }(\mathbf{\hat{d}_{2}}\times \mathbf{\hat{J%
})}  \notag \\
&&\!+(\mathbf{\hat{d}_{2}}\times \mathbf{\hat{S}_{2})(\mathbf{\hat{J}\cdot
\hat{S}_{2})}]}\Bigr\}, \\
\frac{d}{dt}(\mathbf{\hat{J}\cdot \hat{d}_{i})\!} &\mathbf{=}&\mathbf{\!}%
\frac{S_{i}}{\mathcal{I}_{i}}(\mathbf{\hat{S}}_{\mathbf{i}}\times \mathbf{%
\hat{d}_{i})\cdot \hat{J}.}
\end{eqnarray}%
We can compute the scalar products of $\mathbf{\hat{S}}_{\mathbf{i}}\mathbf{%
\cdot \hat{d}_{i}}$ from Eqs. (\ref{DDpure1}) and (\ref{DDpure4}) as%
\begin{equation}
\frac{d}{dt}(\mathbf{\hat{S}}_{\mathbf{i}}\mathbf{\cdot \hat{d}_{i})}=\
\frac{d_{1}d_{2}(\mathbf{\hat{S}}_{\mathbf{i}}\mathbf{\cdot \hat{d}_{i})}}{%
2r^{3}S_{i}}[\mathbf{\hat{d}}_{\mathbf{j}}\mathbf{+}3(\mathbf{\hat{d}}_{%
\mathbf{j}}\mathbf{\cdot \hat{L}}_{\mathbf{N}})\mathbf{\hat{L}}_{\mathbf{N}}]%
\mathbf{\cdot }(\mathbf{\hat{d}_{i}}\times \mathbf{\hat{S}_{i})\mid }_{i\neq
j},
\end{equation}%
where $i,j=1..2$. Assuming that the quadratic terms in magnetic dipoles can
be dropped ($\mathcal{O}(d_{1}d_{2})\approx 0$) we get simple evolution
equations for the pure DD case, where the introduced angles $\cos \beta _{i}=%
\mathbf{\hat{J}\cdot \hat{S}_{i}}$, $\cos \alpha _{i}=\mathbf{\hat{S}%
_{i}\cdot \hat{d}_{i}}$ (see Fig. \ref{spingeometry}) are quasi-constants
\begin{eqnarray}
\frac{d}{dt}(\mathbf{\hat{J}\cdot \hat{S}_{i})} &\approx &0\ ,
\label{simple_DDpure1} \\
\frac{d}{dt}\mathbf{(\hat{S}_{i}\cdot \hat{d}_{i})} &\approx &0\ ,
\label{simple_DDpure2} \\
\frac{d}{dt}(\mathbf{\hat{J}\cdot \hat{d}_{i})} &\mathbf{=}&\frac{S_{i}}{%
\mathcal{I}_{i}}(\mathbf{\hat{S}}_{\mathbf{i}}\times \mathbf{\hat{d}%
_{i})\cdot \hat{J}.}  \label{simple_DDpure3}
\end{eqnarray}%
This system of equations can be easily integrated because from Eqs. (\ref%
{S1mag},\ref{S2mag}) $\dot{S}_{i}=\mathcal{O}(d_{1}d_{2}/r^{3})\approx 0$.
We transform Eq. (\ref{simple_DDpure3}) to
\begin{eqnarray}
\left( \frac{d}{dt}(\mathbf{\hat{J}\cdot \hat{d}_{i})}\right) ^{2} &\mathbf{=%
}&\Omega _{i}^{2}\Bigl[1-(\mathbf{\hat{J}\cdot \hat{S}_{i})}^{2}-\mathbf{(%
\hat{S}_{i}\cdot \hat{d}_{i})}^{2}-(\mathbf{\hat{J}\cdot \hat{d}_{i})}^{2}
\notag \\
&&+2(\mathbf{\hat{J}\cdot \hat{S}_{i})(\hat{S}_{i}\cdot \hat{d}_{i})}(%
\mathbf{\hat{J}\cdot \hat{d}_{i})}\Bigr]^{2},  \label{simple_DDpure}
\end{eqnarray}%
where $\Omega _{i}\mathbf{=}S_{i}/\mathcal{I}_{i}$ are the spin
frequencies or rotational angular velocities of NSs (e.g., see Table \ref%
{table1}). By introducing the following relative angles of $\cos \delta _{i}=%
\mathbf{\hat{J}\cdot \hat{d}_{i}}$, $\cos \beta _{i0}=(\mathbf{\hat{J}\cdot
\hat{S}}_{\mathbf{i}})_{0}$ and $\cos \alpha _{i0}=(\mathbf{\hat{S}}_{%
\mathbf{i}}\mathbf{\cdot \hat{d}_{i})}_{0}$, where angles $\alpha _{i0}$ and
$\kappa _{i0}$ are quasi-constants, we get a harmonic oscillator equation in
Eq. (\ref{simple_DDpure})%
\begin{equation}
\frac{d^{2}\cos \delta _{i}}{dt^{2}}\mathbf{+}\Omega _{i}^{2}\cos \delta
_{i}=C_{i},  \label{harm_osci}
\end{equation}%
with an introduced quasi-constant of $C_{i}=\Omega _{i}^{2}\cos \beta
_{i0}\cos \alpha _{i0}$. Thus, the general solution of Eq. (\ref{harm_osci})
is%
\begin{equation}
\cos \delta _{i}=C_{i1}\cos \Omega _{i}t+C_{i2}\sin \Omega _{i}t+\frac{C_{i}%
}{\Omega _{i}^{2}}.  \label{gen_sol}
\end{equation}%
If we choose initial values of $\cos \delta _{i}(0)=\cos \delta _{i0}$ and $%
\frac{d}{dt}\cos \delta _{i}(0)=0$ then $C_{i1}=\cos \delta _{i0}-\cos \beta
_{i0}\cos \alpha _{i0}$ and $C_{i2}=0$. Then we get a particular solution
for Eq. (\ref{gen_sol})
\begin{eqnarray}
\cos \delta _{i} &=&\left( \cos \delta _{i0}-\cos \beta _{i0}\cos \alpha
_{i0}\right) \cos \Omega _{i}t  \notag \\
&&+\cos \beta _{i0}\cos \alpha _{i0},  \label{DDpure_sol}
\end{eqnarray}%
with the initial value notations of $\cos \delta _{i0}=(\mathbf{\hat{J}\cdot
\hat{d}_{i})}_{0}$, $\cos \beta _{i0}=(\mathbf{\hat{J}\cdot \hat{S}}_{%
\mathbf{i}})_{0}$ and $\cos \alpha _{i0}=(\mathbf{\hat{S}}_{\mathbf{i}}%
\mathbf{\cdot \hat{d}_{i}})_{0}$. We found that the difference between the
perturbative and numerical solutions is small with a maximum difference in
amplitude of $2\%$ even for large magnetic dipole parameters of $0.005$ with
spin parameters of $0.001$ (see, Figs. \ref{evol0} and \ref{evol02}).

Similar equations can be derived for the orbital angular momentum from Eqs. (%
\ref{DDpure3},\ref{DDpure4})
\begin{equation}
\frac{d}{dt}(\mathbf{\hat{L}}_{\mathbf{N}}\mathbf{\cdot \hat{d}_{i})}\mathbf{%
=}\frac{S_{i}}{\mathcal{I}_{i}}(\mathbf{\hat{S}}_{\mathbf{i}}\times \mathbf{%
\hat{d}_{i})\cdot \mathbf{\hat{L}}_{\mathbf{N}},}  \label{simple_DDpure_L}
\end{equation}%
which also leads to an oscillator equation because $\frac{d}{dt}(\mathbf{%
\hat{L}}_{\mathbf{N}}\mathbf{\cdot \hat{S}_{i})}\approx 0\ $according to the
previous approximation
\begin{equation}
\frac{d^{2}\cos \varrho _{i}}{dt^{2}}\mathbf{+}\Omega _{i}^{2}\cos \varrho
_{i}=B_{i},  \label{harm_osci_L}
\end{equation}%
where $\cos \varrho _{i}=\mathbf{\hat{L}}_{\mathbf{N}}\mathbf{\cdot \hat{d}%
_{i}}$ and $B_{i}=\Omega _{i}^{2}\cos \kappa _{i0}\cos \alpha _{i0}$ with $%
\cos \kappa _{i0}=(\mathbf{\hat{L}}_{\mathbf{N}}\mathbf{\cdot \hat{S}}_{%
\mathbf{i}})_{0}$. Then the solution for the initial values of\ $\cos
\varrho _{i}(0)=(\mathbf{\hat{L}}_{\mathbf{N}}\mathbf{\cdot \hat{d}_{i})}%
_{0}\equiv \cos \varrho _{i0}$ and $\frac{d}{dt}\cos \varrho _{i}(0)=0$ \ is%
\begin{eqnarray}
\cos \varrho _{i} &=&\left( \cos \varrho _{i0}-\cos \kappa _{i0}\cos \alpha
_{i0}\right) \cos \Omega _{i}t  \notag \\
&&+\cos \kappa _{i0}\cos \alpha _{i0}.  \label{DDpure_sol_L}
\end{eqnarray}

\subsubsection{Equal mass moments of inertia case}

It can be noted that using Eq. (\ref{ddot}) in Eqs. (\ref{S1mag},\ref{S2mag}%
) we get

\begin{eqnarray}
\frac{d}{dt}S_{1}^{2} &=&-\frac{\mathcal{I}_{1}}{r^{3}}\left[ \mathbf{%
\mathbf{d}_{2}+}3(\mathbf{d_{2}\cdot \hat{L}}_{\mathbf{N}})\mathbf{\hat{L}}_{%
\mathbf{N}}\right] \mathbf{\cdot \dot{d}_{1}},  \label{spinmag1} \\
\frac{d}{dt}S_{2}^{2} &=&-\frac{\mathcal{I}_{2}}{r^{3}}\left[ \mathbf{%
\mathbf{d}_{1}+}3(\mathbf{d_{1}\cdot \hat{L}}_{\mathbf{N}})\mathbf{\hat{L}}_{%
\mathbf{N}}\right] \mathbf{\cdot \dot{d}_{2}}.  \label{spinmag2}
\end{eqnarray}%
If we assume equal mass moments of inertia $\mathcal{I}_{1}=\mathcal{I}%
_{2}\equiv \mathcal{I}$ then the summation of Eqs. (\ref{spinmag1},\ref%
{spinmag2}) leads to a total time derivative, as
\begin{equation}
\frac{d}{dt}\left( S_{1}^{2}+S_{2}^{2}\right) =-\frac{\mathcal{I}}{r^{3}}%
\frac{d}{dt}[\mathbf{d_{1}\cdot \mathbf{d}_{2}\ +}3(\mathbf{d_{1}\cdot \hat{L%
}}_{\mathbf{N}})(\mathbf{\mathbf{d}_{2}\cdot \hat{L}}_{\mathbf{N}})].
\label{totalD}
\end{equation}%
Integrating Eq. (\ref{totalD}) and using that $d_{i}$ are conserved
quantities we get%
\begin{eqnarray}
S_{1}^{2}+S_{2}^{2} &=&-\frac{\mathcal{I}d_{1}d_{2}}{r^{3}}[\mathbf{\hat{d}%
_{1}\cdot \hat{d}_{2}+}3(\mathbf{\hat{d}_{1}\cdot \hat{L}}_{\mathbf{N}})(%
\mathbf{\hat{d}_{2}\cdot \hat{L}}_{\mathbf{N}})]+C_{0},  \notag \\
C_{0} &=&S_{10}^{2}+S_{20}^{2}  \notag \\
&&-\frac{\mathcal{I}d_{1}d_{2}}{r^{3}}[\mathbf{\hat{d}_{1}\cdot \hat{d}_{2}+}%
3(\mathbf{\hat{d}_{1}\cdot \hat{L}}_{\mathbf{N}})(\mathbf{\hat{d}_{2}\cdot
\hat{L}}_{\mathbf{N}})]_{0},  \label{exacts}
\end{eqnarray}%
where $S_{i0}$ are the initial magnitudes of spin vectors $S_{i}$ and $\left[
\mathbf{...}\right] _{0}$ contains the relative angles between the initial
magnetic dipole moments and the Newtonian orbital angular momentum at $t=0$.
Thus, we obtain an exact relation between the magnitudes of spins and the
geometry of the magnetic dipole moments for the equal mass moments of
inertia case.

\begin{widetext}

\begin{figure}[tbh]
\begin{center}
\includegraphics[width=0.32
\textwidth]{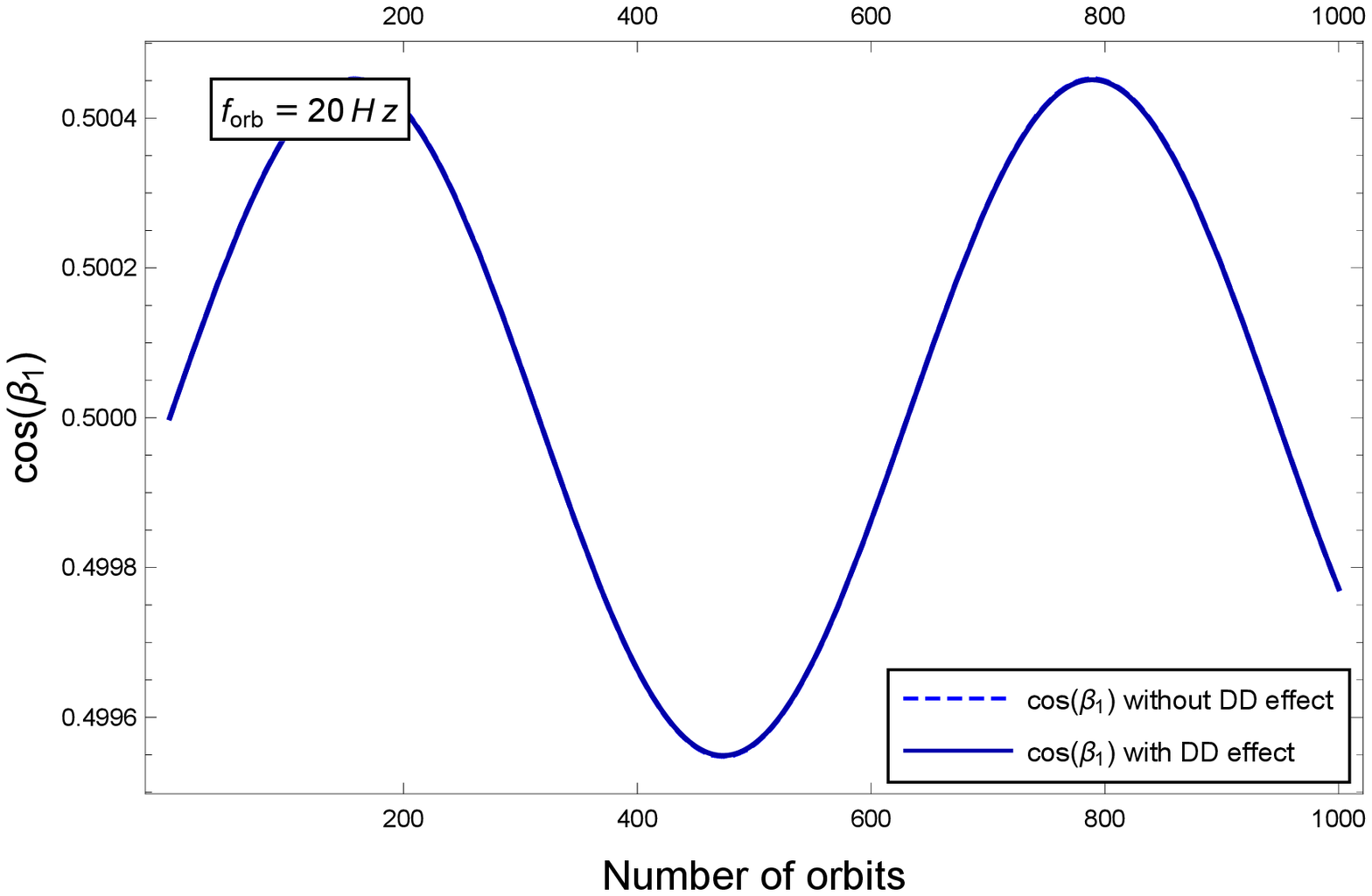} \includegraphics[width=0.32%
\textwidth]{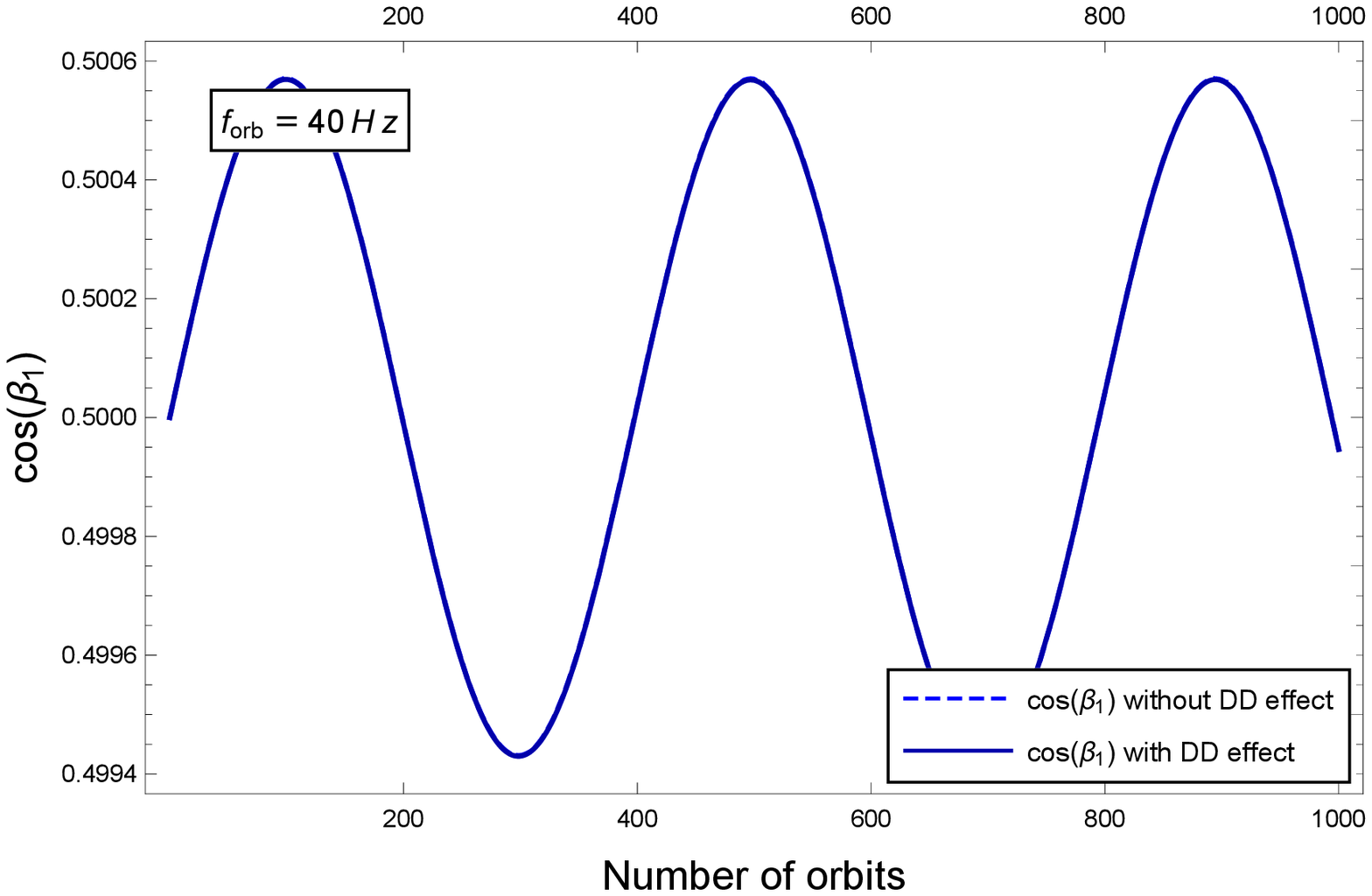} \includegraphics[width=0.32%
\textwidth]{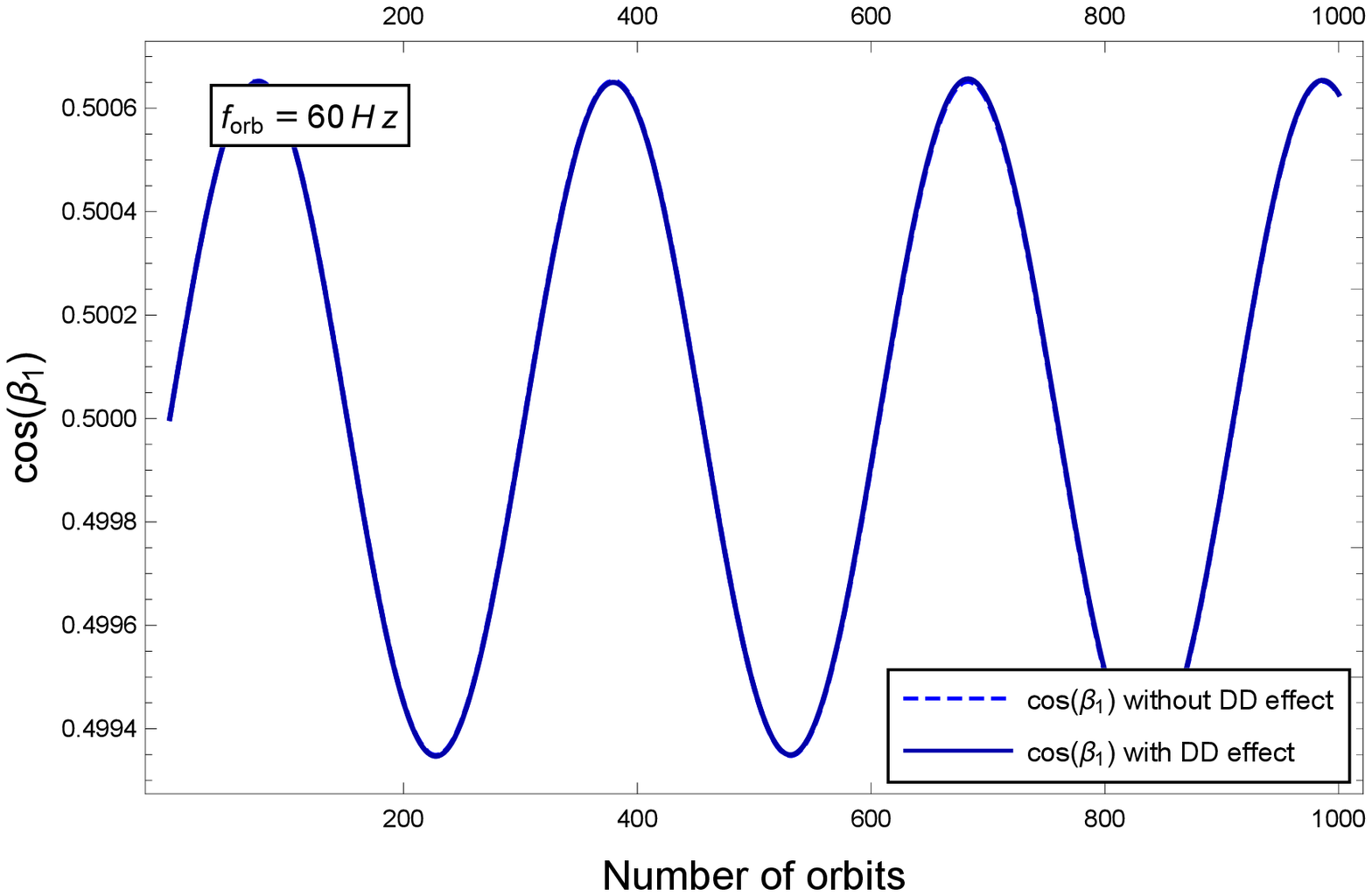} \includegraphics[width=0.32%
\textwidth]{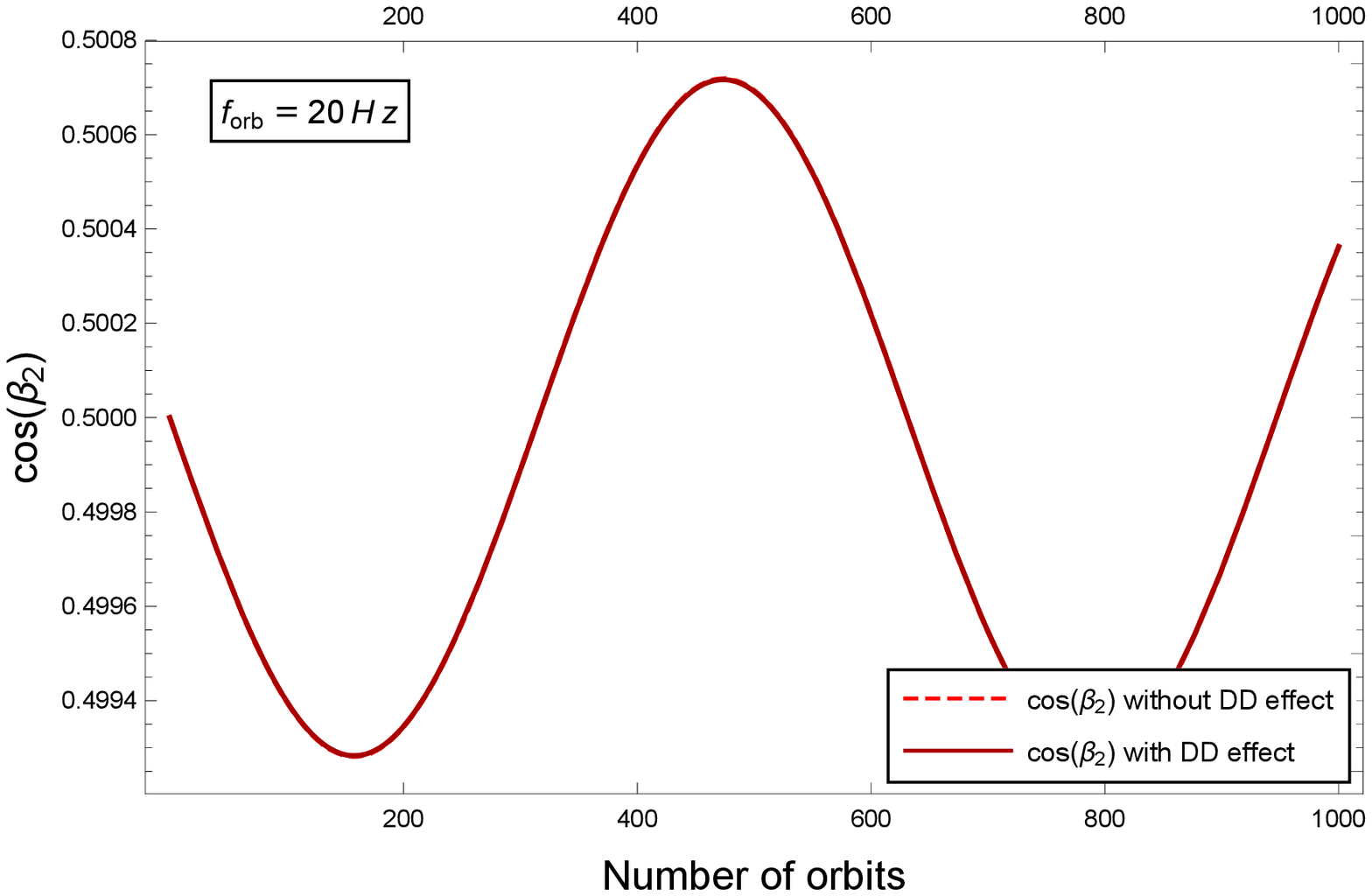} \includegraphics[width=0.32%
\textwidth]{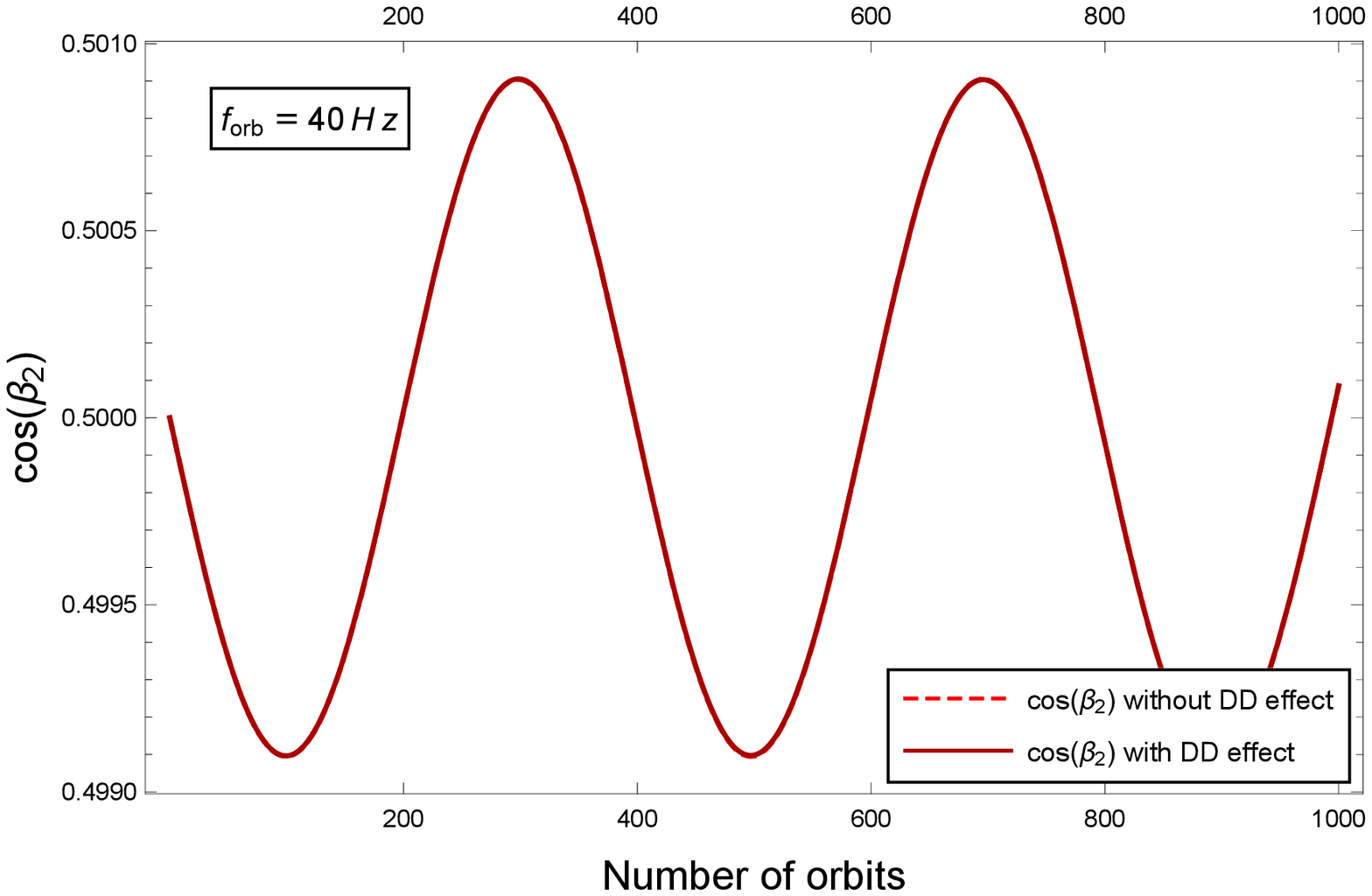} \includegraphics[width=0.32%
\textwidth]{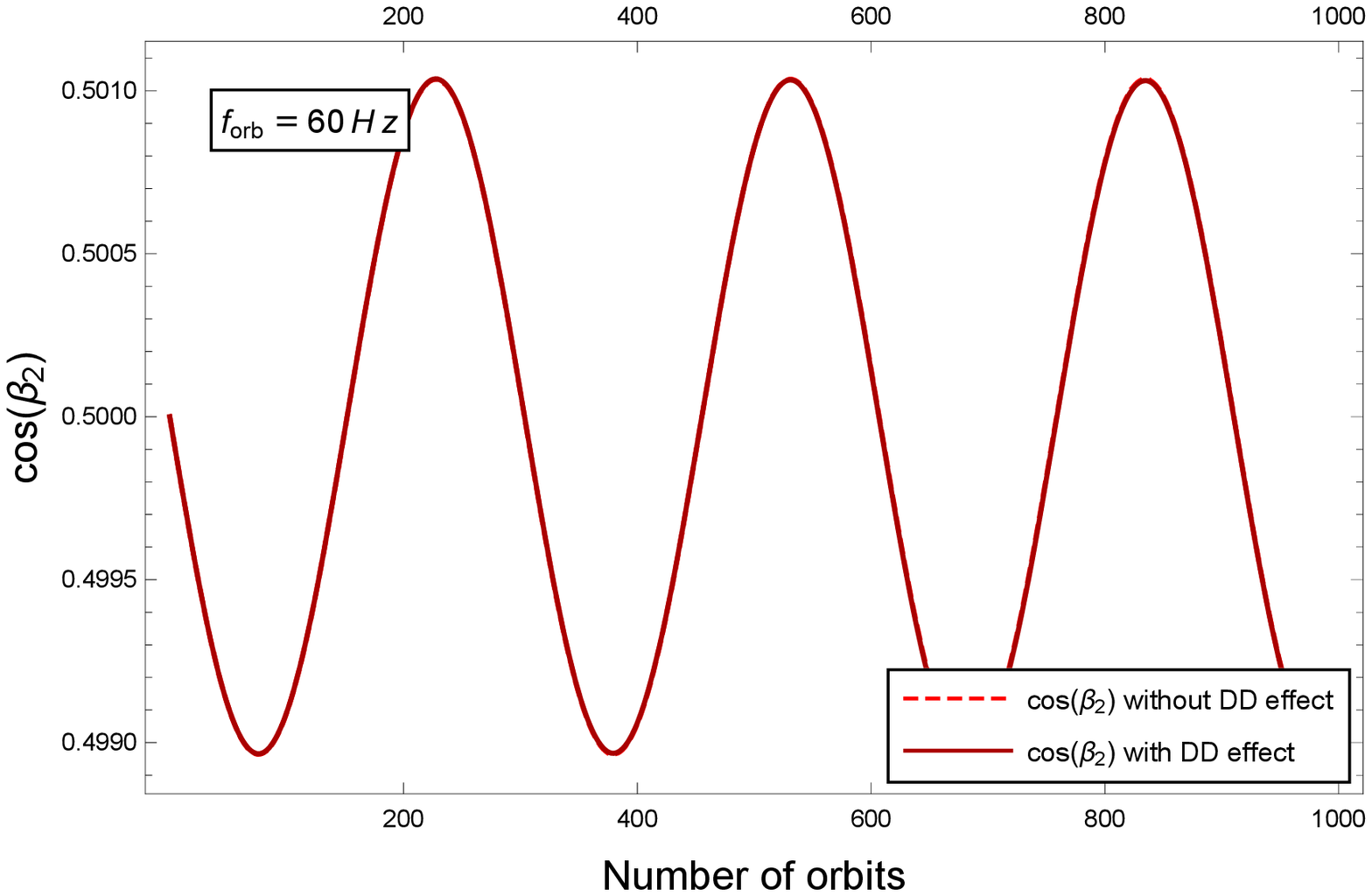}
\end{center}
\caption{Evolutions of relative angles $\cos \protect\beta _{i}=\mathbf{\hat{%
J}\cdot \hat{S}}_{\mathbf{i}}$. Masses of the binary NS are
$1.4M_{\odot }$ and $1.2M_{\odot }$. Orbital frequencies are chosen
from left to right as $20
$, $40$ and $60$ Hz, respectively. The dimensionless spin parameter is $%
\protect\chi =0.001$ and the ratio parameter is $\Delta =0.1$ ($=\protect\chi ^{d}/%
\protect\chi $). Initial spin and dipole angles in $\mathcal{K}$
are chosen as $\protect\theta _{1}(0)=\protect\pi /4=\protect\theta _{2}(0)$%
, $\protect\varphi _{1}(0)=0$, $\protect\varphi _{2}(0)=\protect\pi /2$, $%
\protect\theta _{1}^{d}(0)=\protect\theta _{1}(0)+\protect\pi /12=\protect%
\theta _{2}^{d}(0)$, $\protect\varphi _{1}^{d}(0)=\protect\varphi
_{1}(0)$ and $\protect\varphi _{2}^{d}(0)=\protect\varphi _{2}(0)$.
It can be seen that the evolutions are harmonics and DD effect can
be neglected in all examples. } \label{evol1}
\end{figure}

\begin{figure}[tbh]
\begin{center}
\includegraphics[width=0.32
\textwidth]{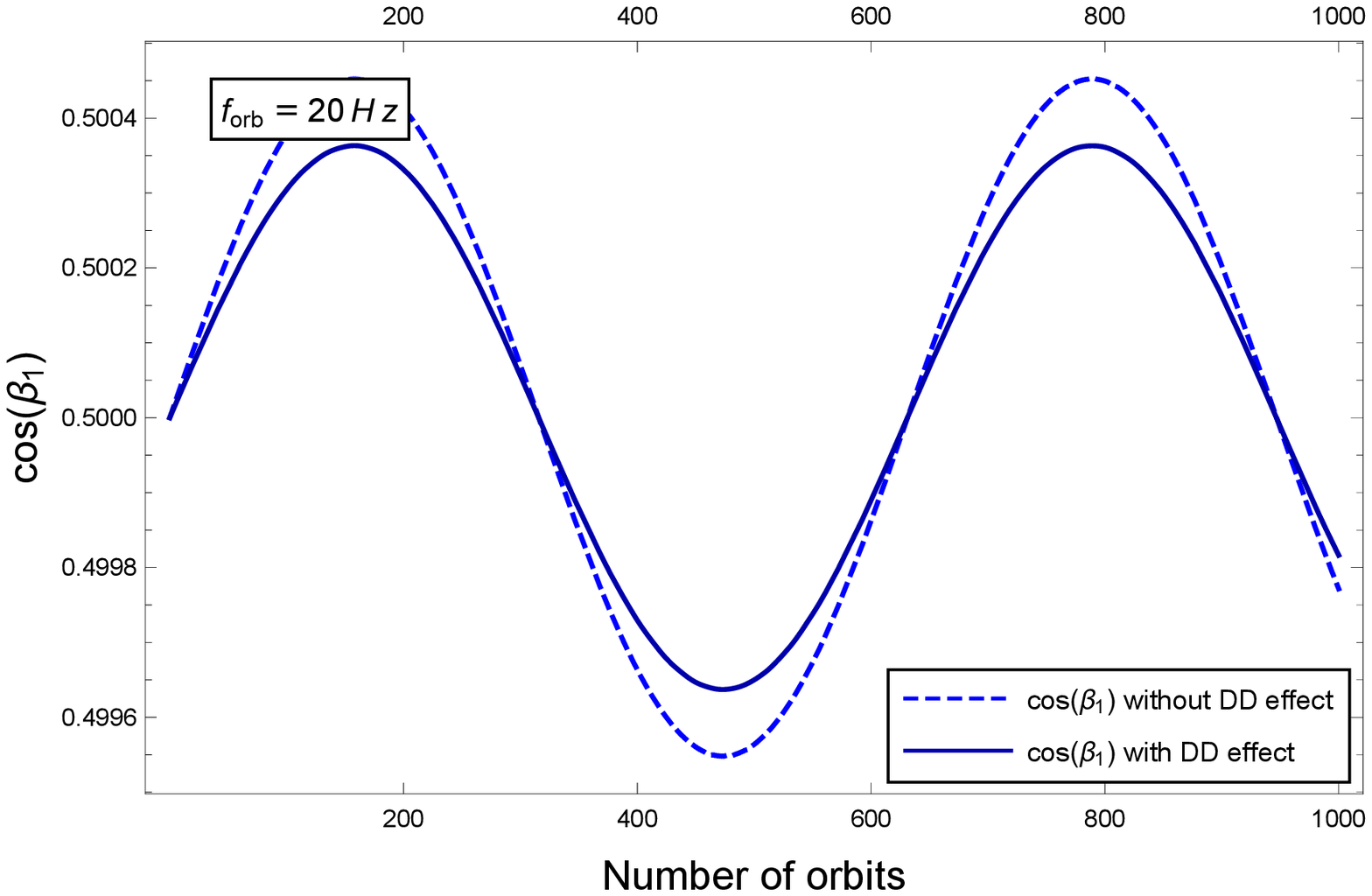} \includegraphics[width=0.32%
\textwidth]{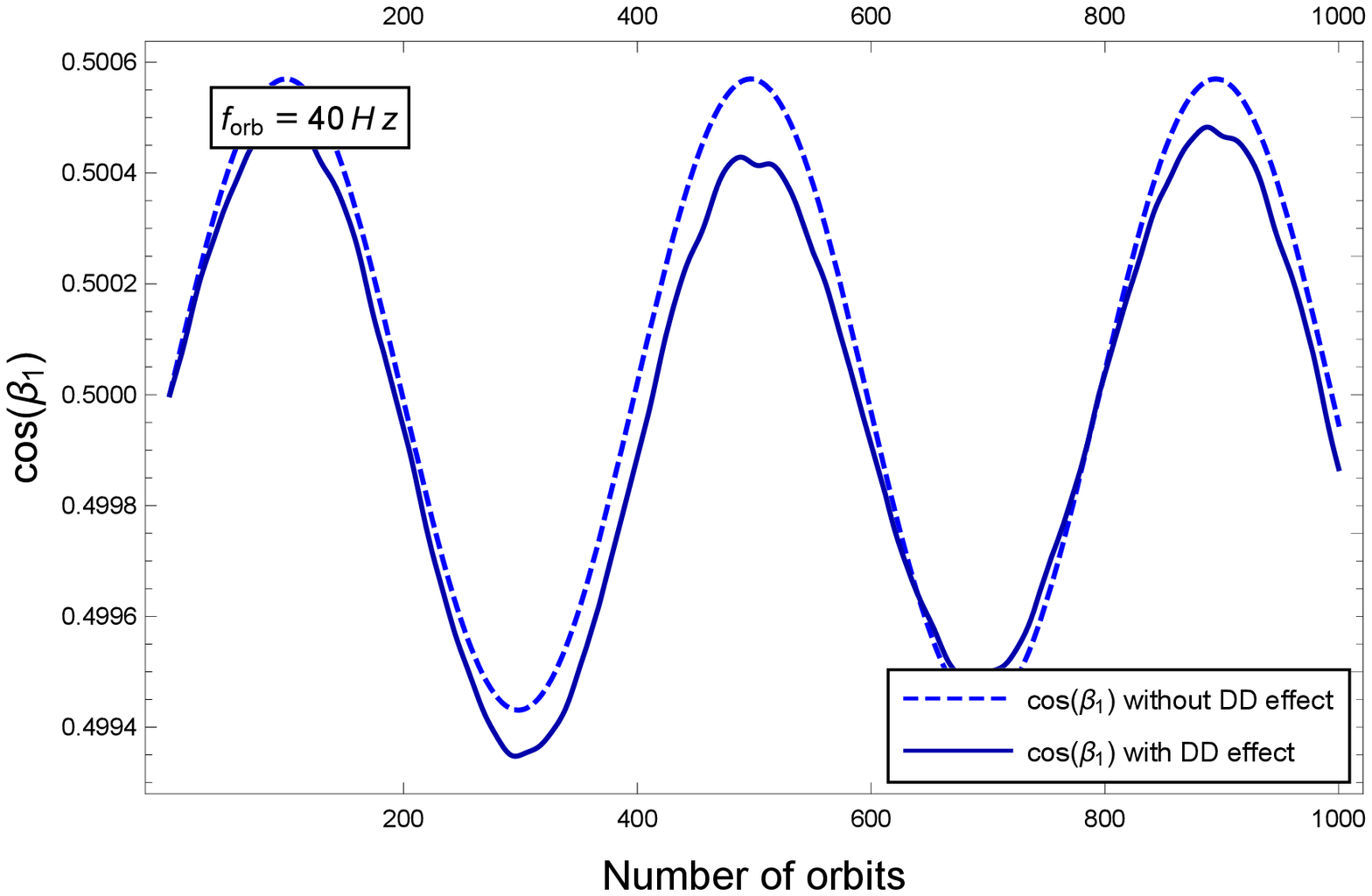} \includegraphics[width=0.32%
\textwidth]{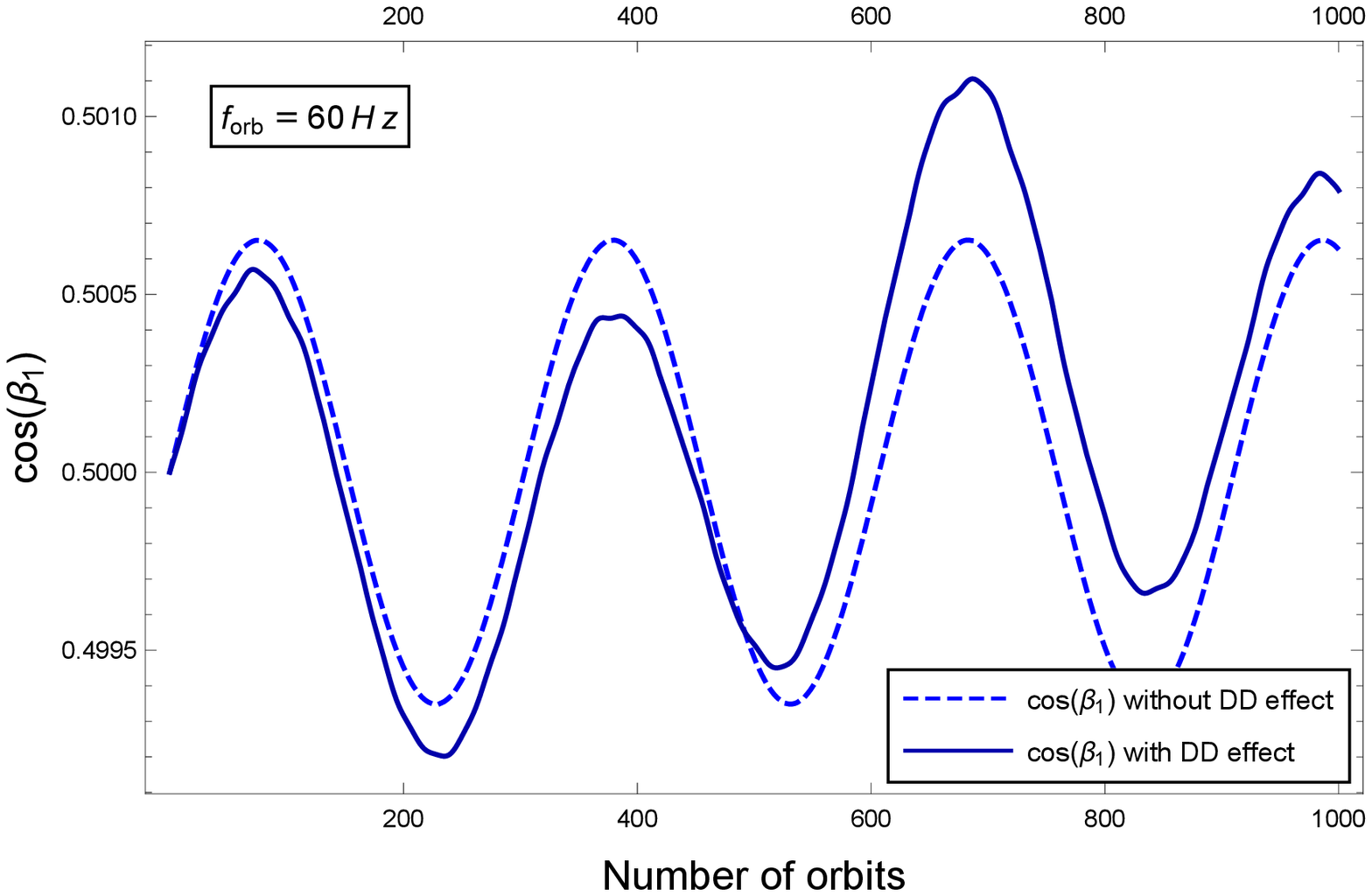} \includegraphics[width=0.32%
\textwidth]{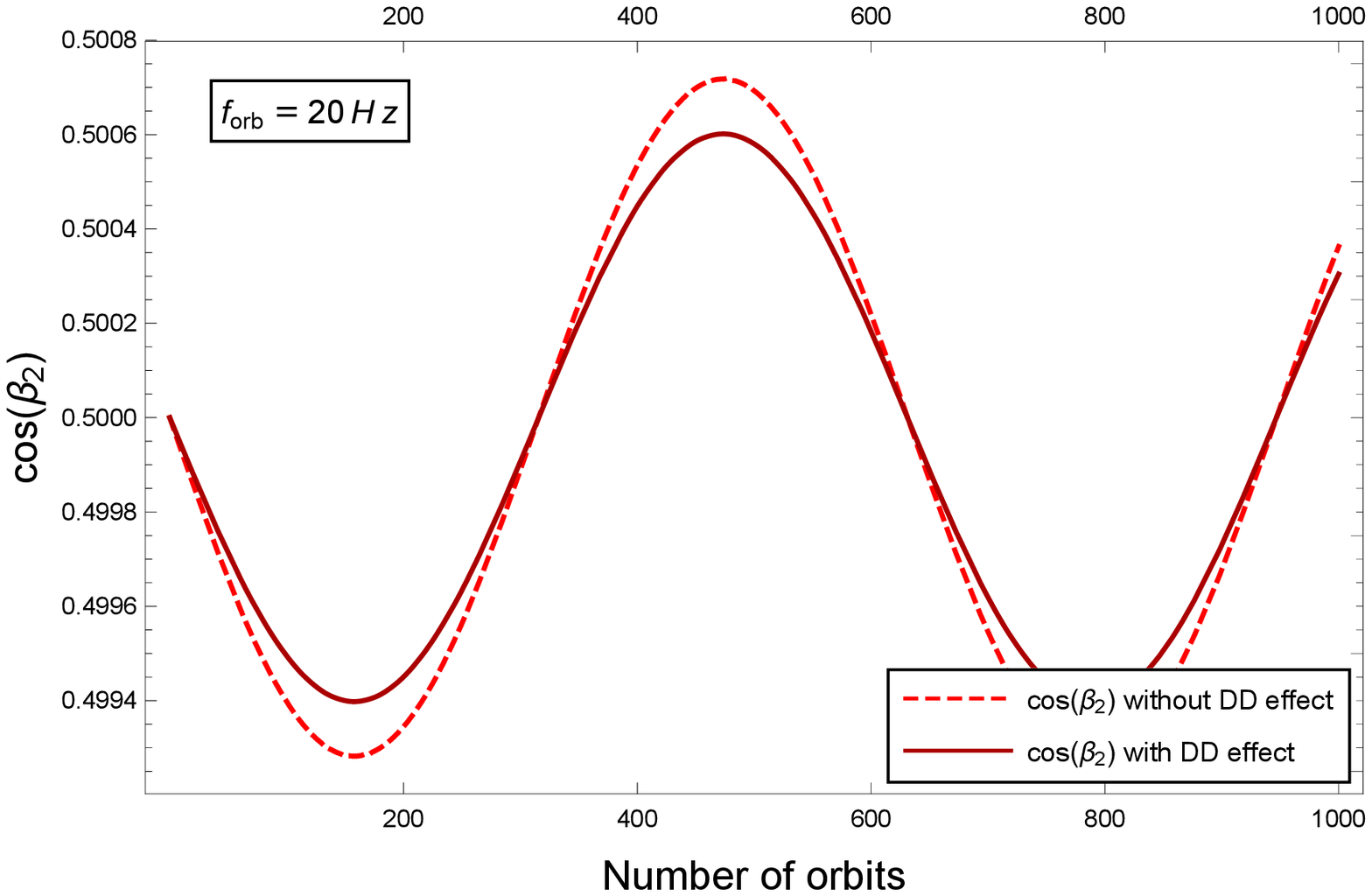} \includegraphics[width=0.32%
\textwidth]{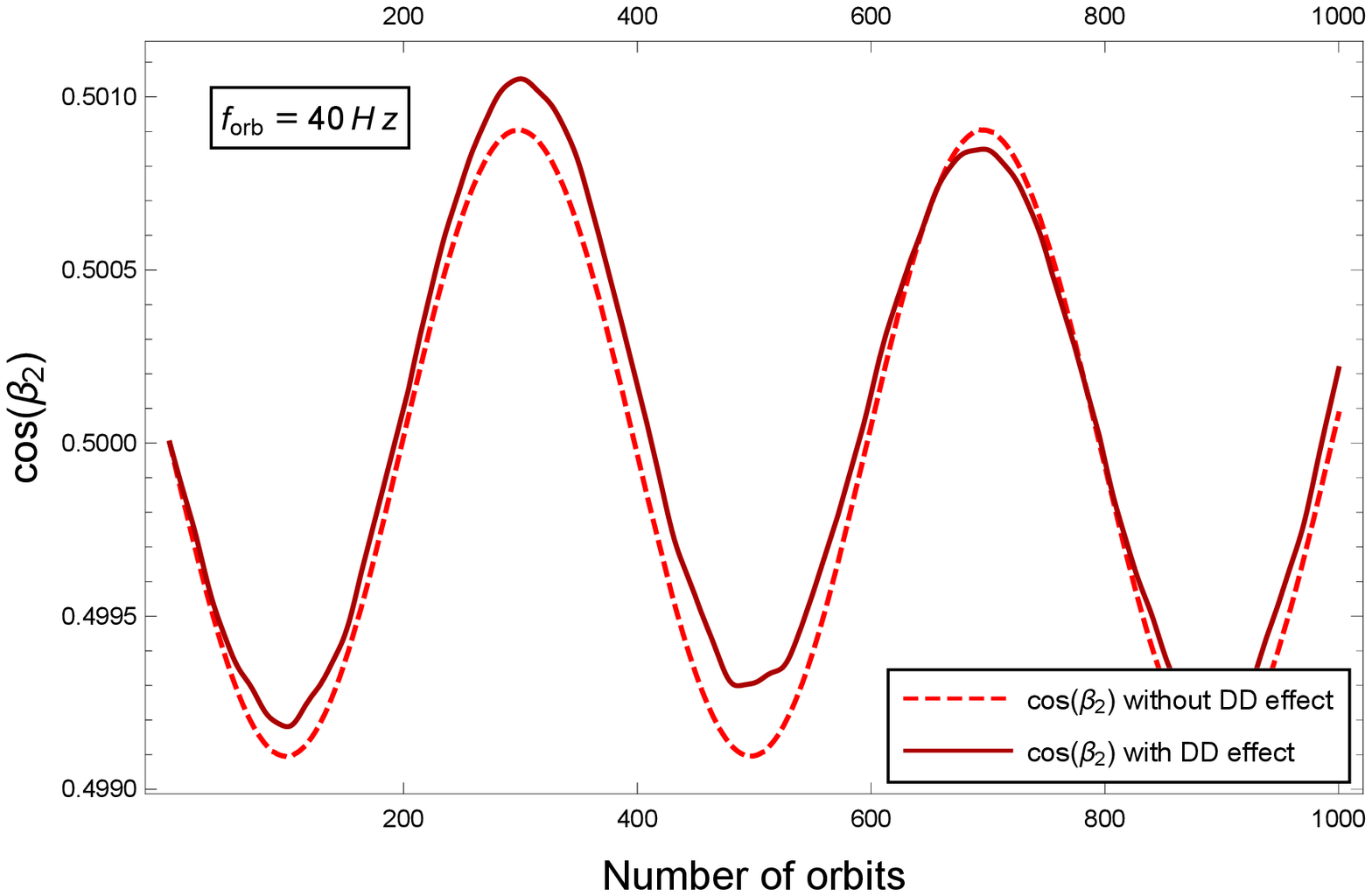} \includegraphics[width=0.32%
\textwidth]{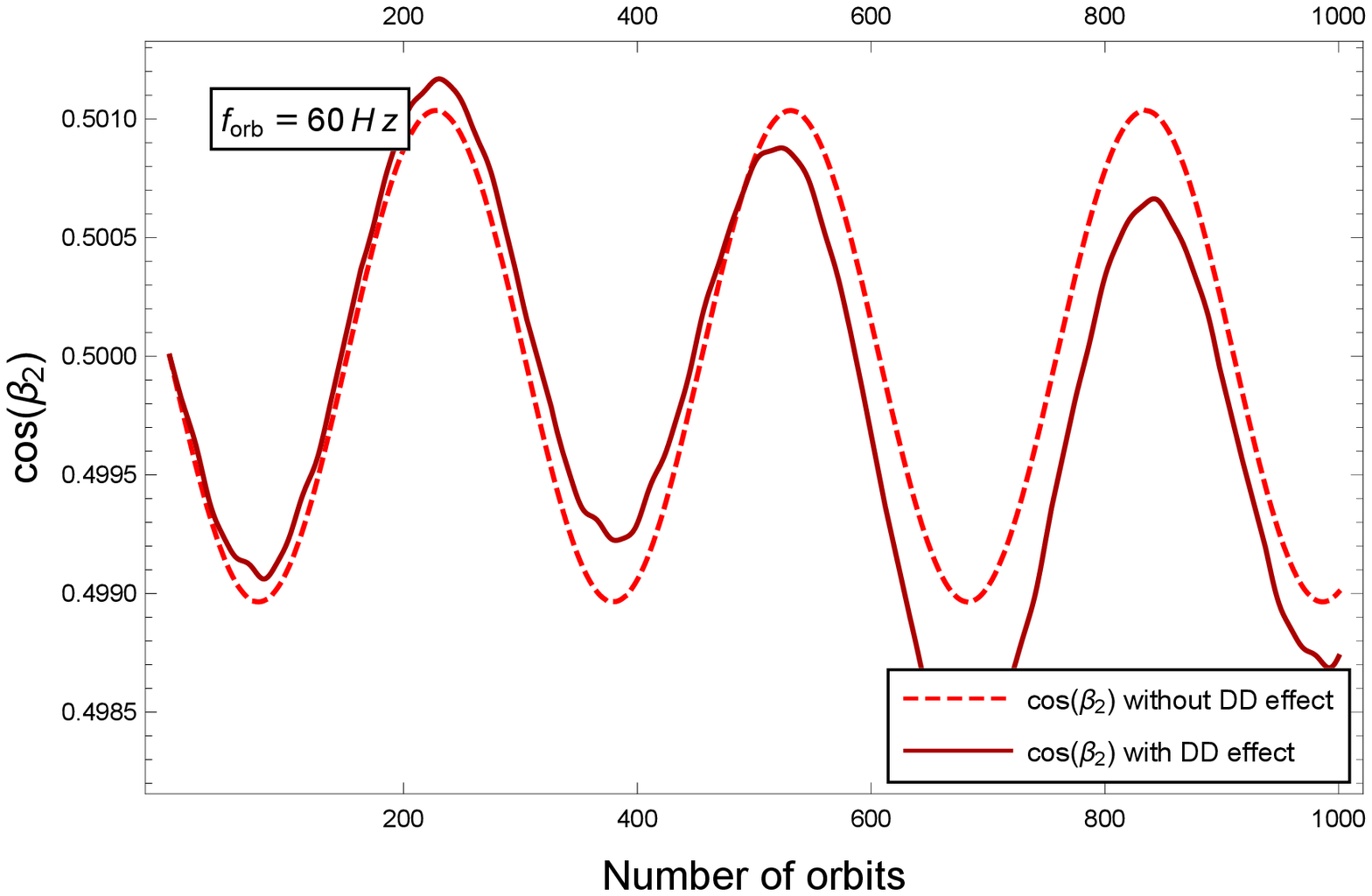}
\end{center}
\caption{Evolutions of relative angles $\cos \protect\beta _{i}$.
The parameters and initial values are the same as in Fig.
\protect\ref{evol1} with the exception of the ratio parameter which
is $\Delta =1$ in this case. The DD effect can be seen appearing,
but the difference in amplitude is rather small, $\symbol{126}1\%$.}
\label{evol2}
\end{figure}

\begin{figure}[tbh]
\begin{center}
\includegraphics[width=0.32
\textwidth]{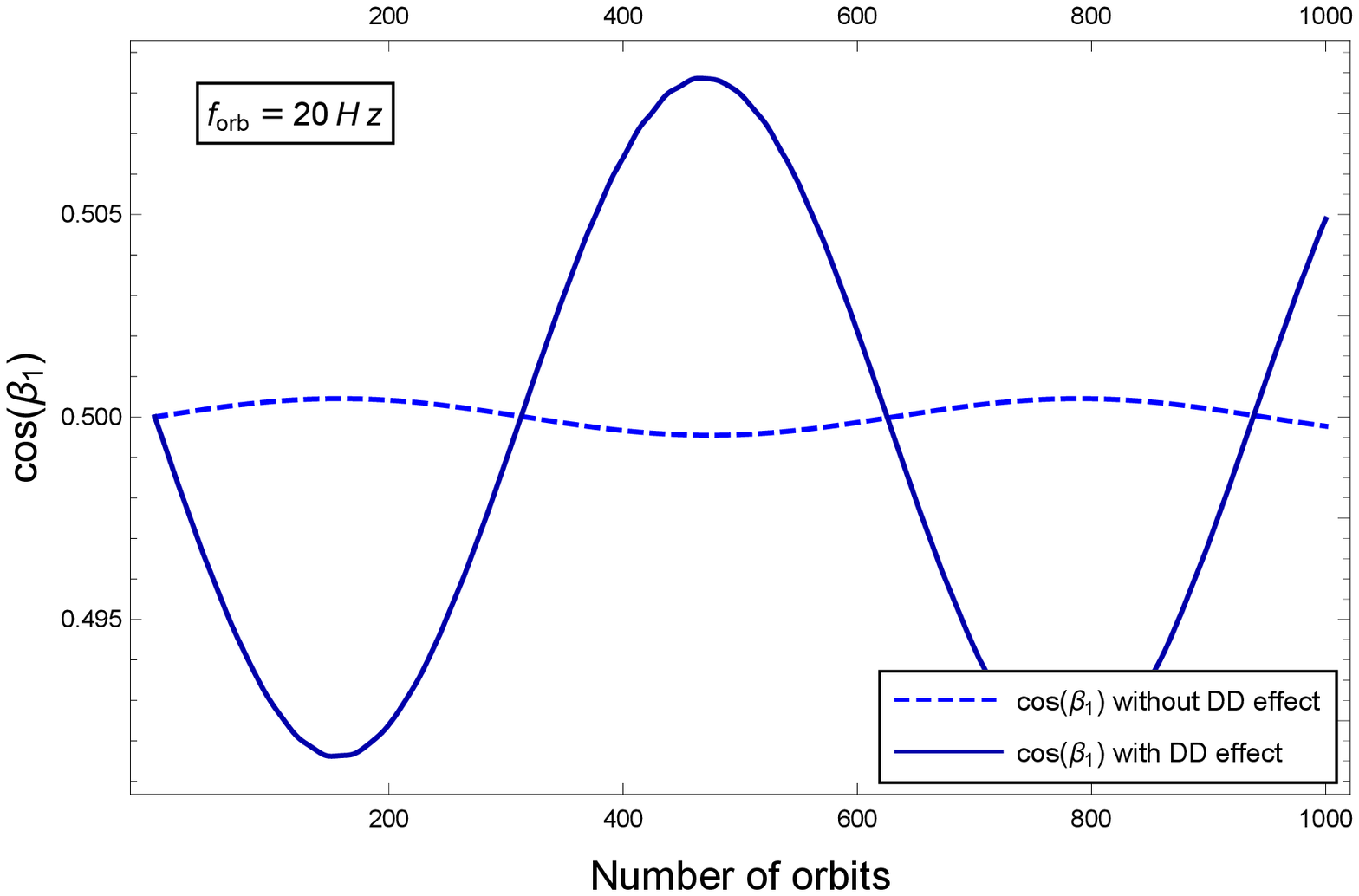} \includegraphics[width=0.32%
\textwidth]{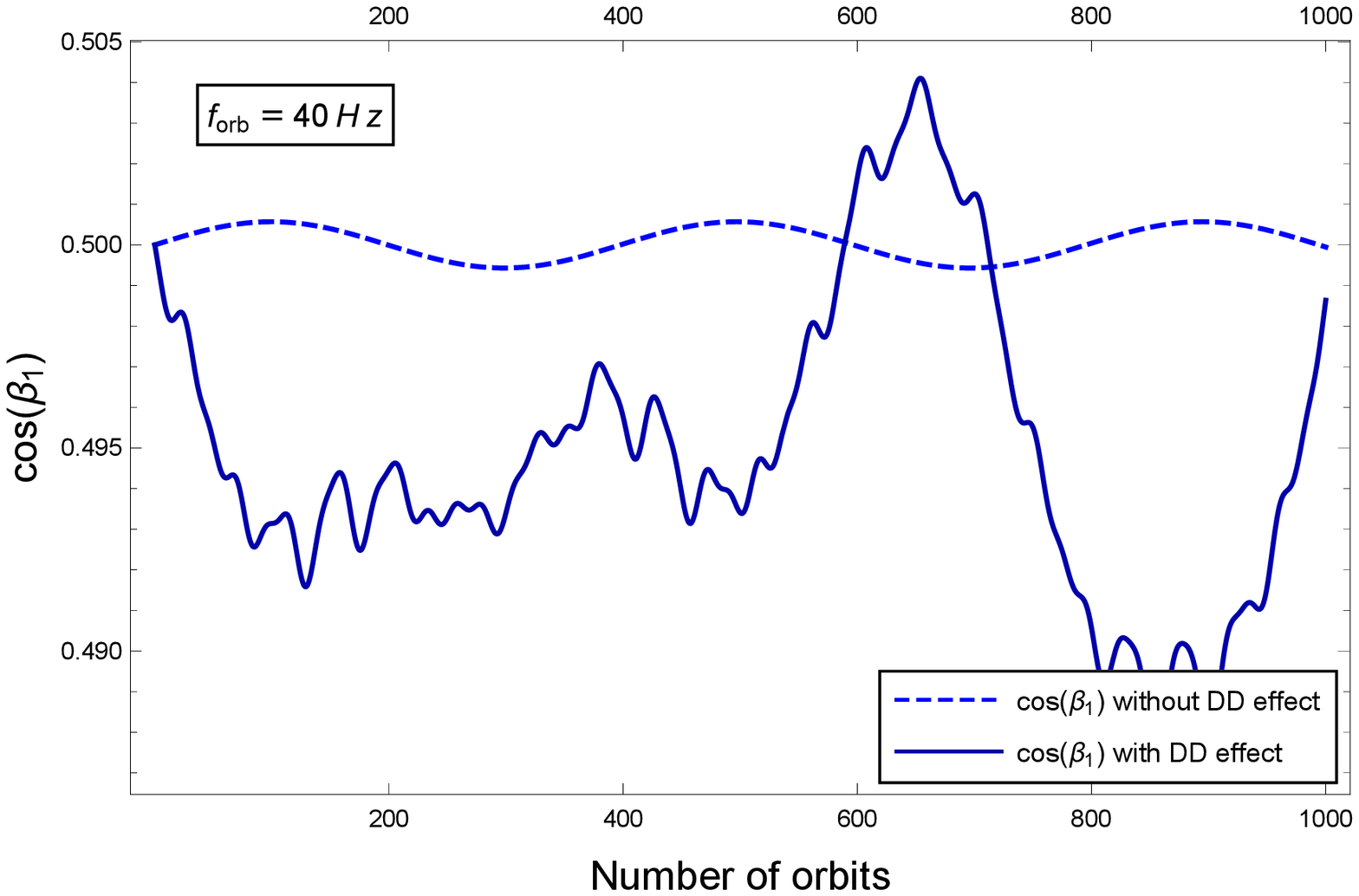} \includegraphics[width=0.32%
\textwidth]{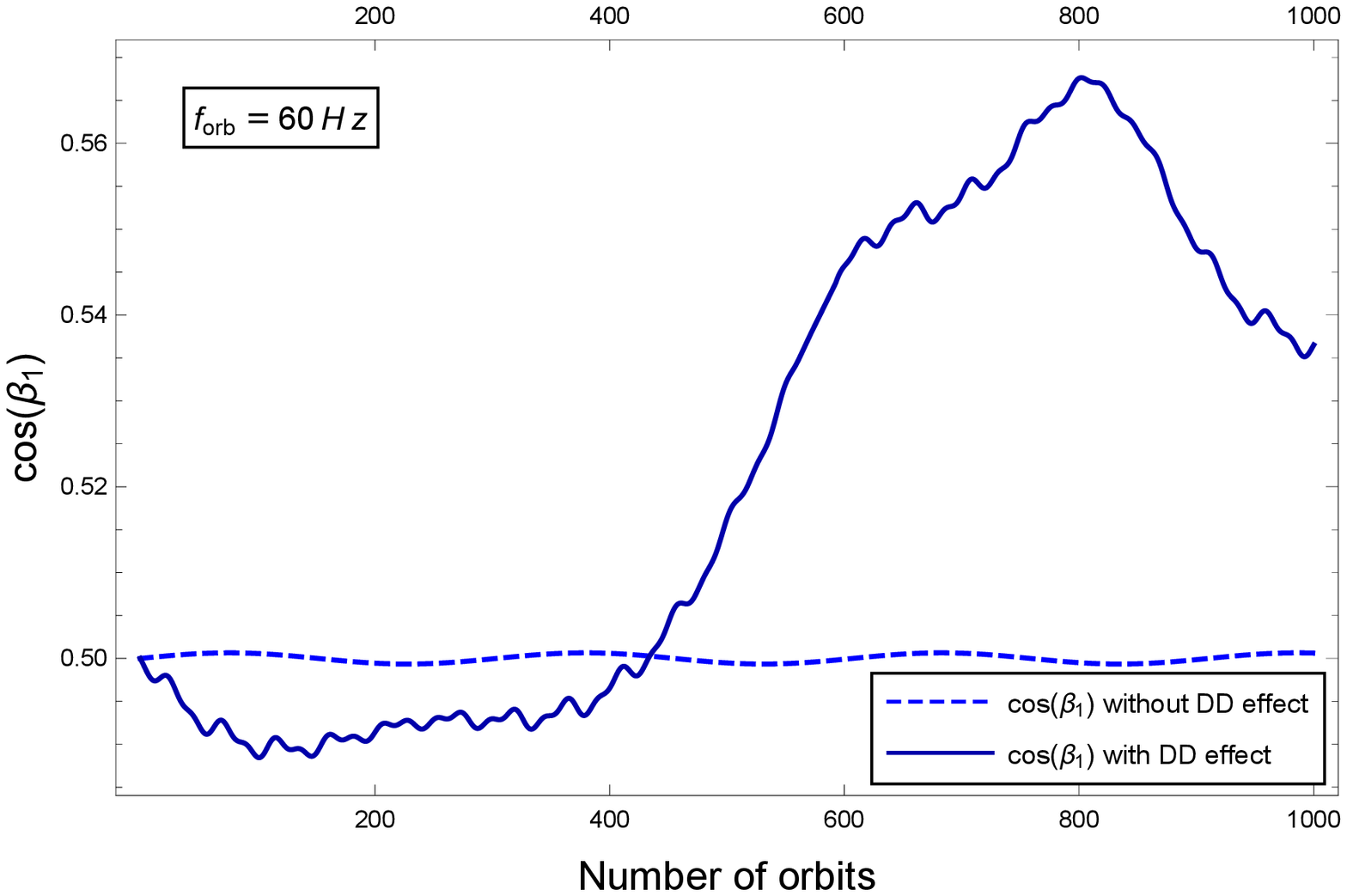} \includegraphics[width=0.32%
\textwidth]{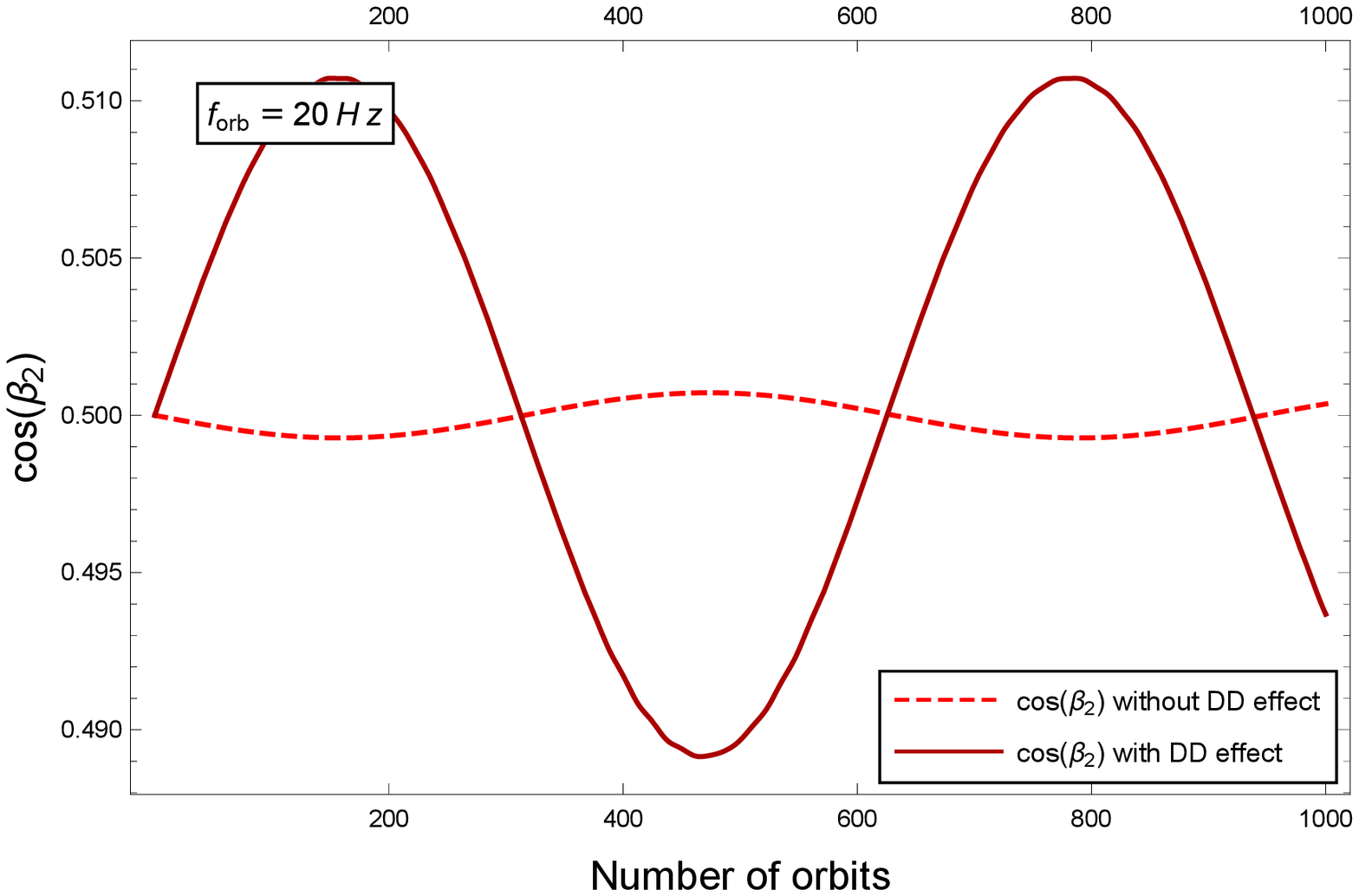} \includegraphics[width=0.32%
\textwidth]{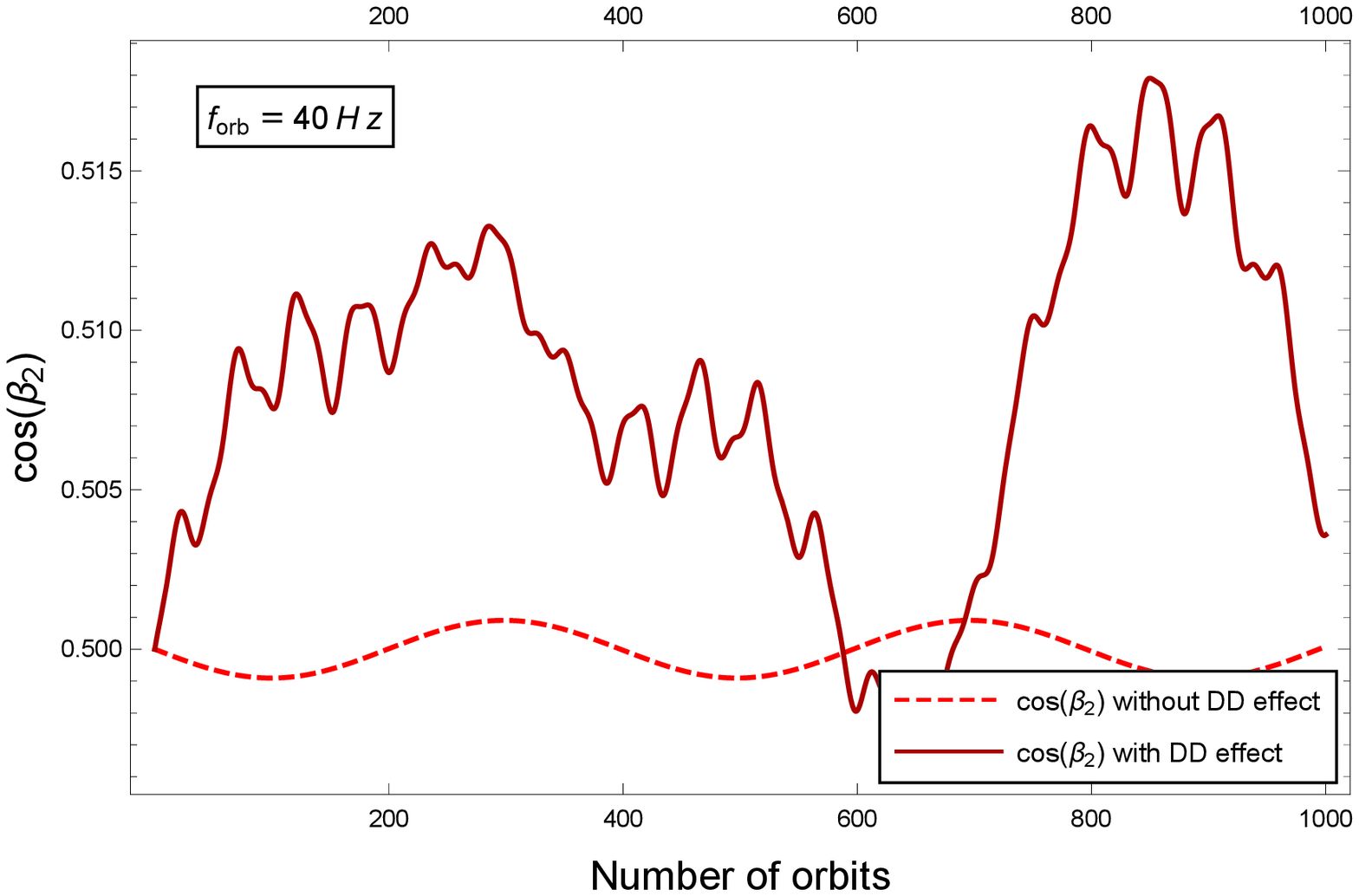} \includegraphics[width=0.32%
\textwidth]{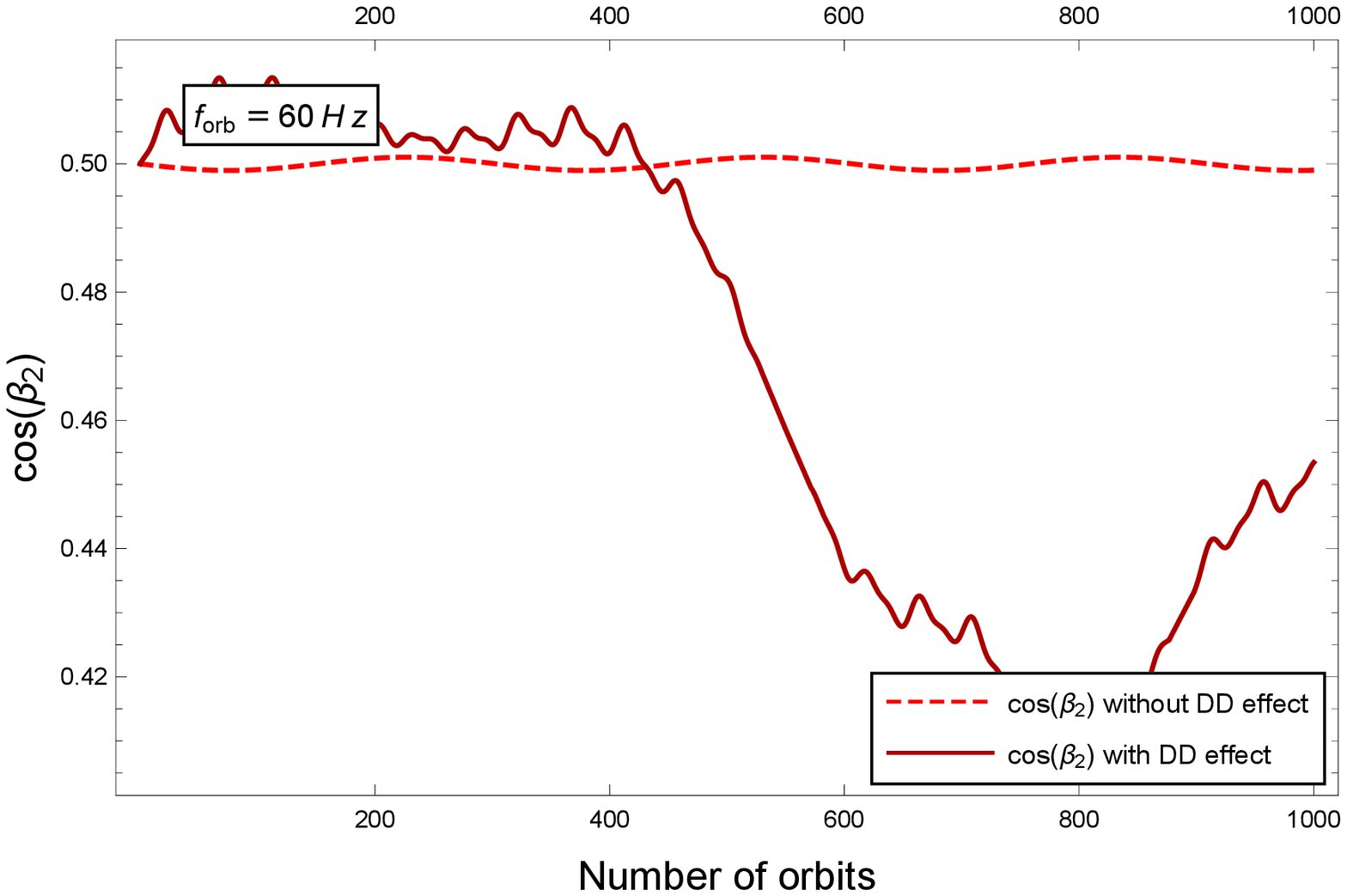}
\end{center}
\caption{Evolutions of relative angles $\cos \protect\beta _{i}$.
The parameters and initial values are the same as in Fig.
\protect\ref{evol1} with the exception of the ratio parameter which
is $\Delta =10.$ It can be seen that the DD effect can be
significant and the difference in amplitude can reach $20\%$.}
\label{evol3}
\end{figure}

\end{widetext}

\begin{figure}[tbh]
\begin{center}
\includegraphics[width=0.45\textwidth]{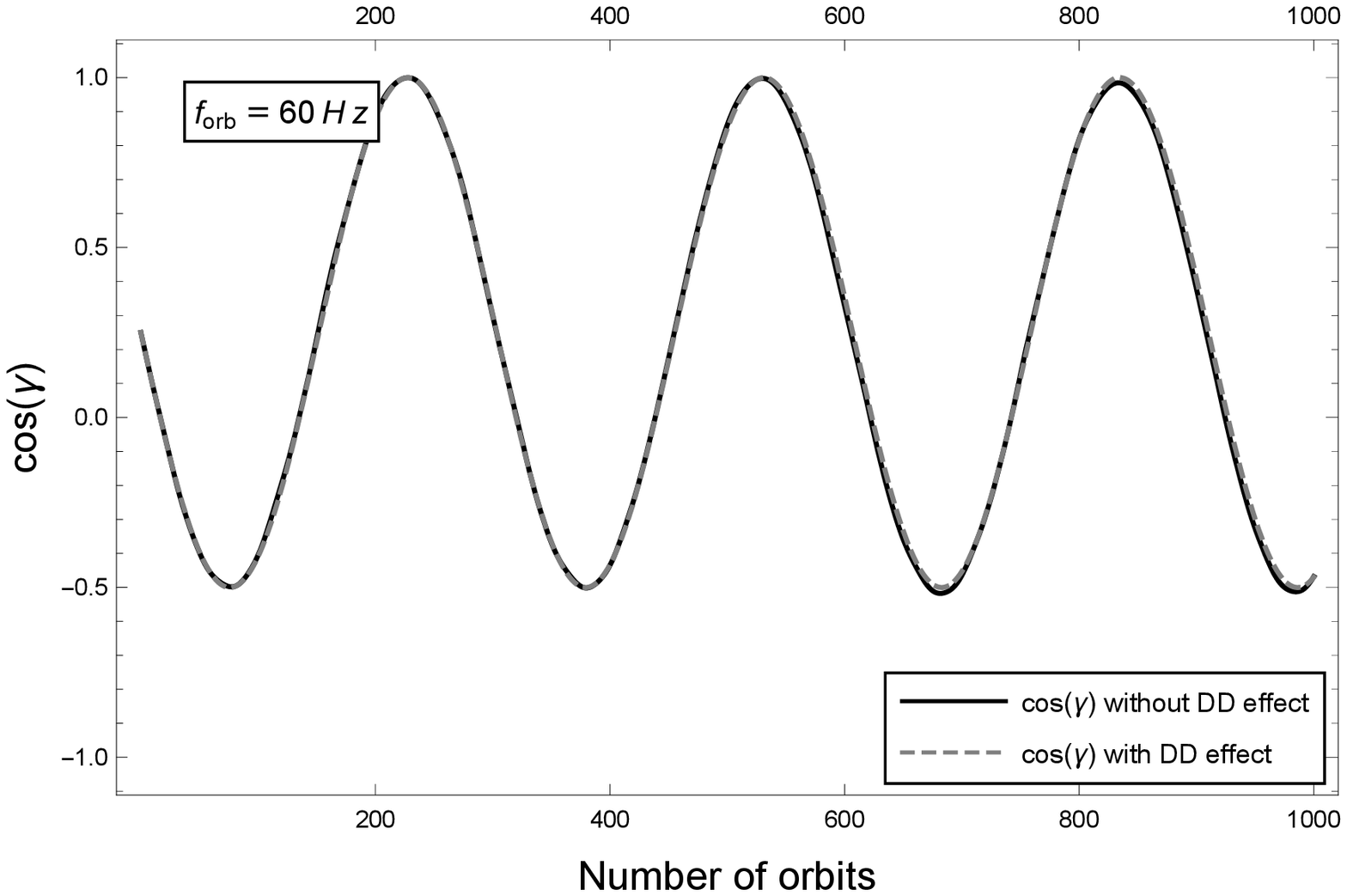}
\end{center}
\caption{Evolutions of relative angle $\cos \protect\gamma $. The
parameters and initial values are the same as in Fig.
\protect\ref{evol3} at orbital frequency of $60$ Hz ($\Delta =10$).}
\label{evol4}
\end{figure}

\section{Numerical analysis for general case}

The coupled system of first order nonlinear differential equations including
the SO, SS, QM and DD terms is very complicated in Eqs. (\ref{S1dot}-\ref%
{ddot}). Therefore to solve this system of equations we analyze it
numerically. We will be using the dimensionless spin parameters $%
\chi _{i}$ and we introduce the dimensionless magnetic dipole
parameters $\chi _{i}^{d}$, which are%
\begin{equation}
\chi _{i}=\frac{cS_{i}}{Gm_{i}^{2}},\qquad \chi _{i}^{d}=\frac{\mu
_{0}^{1/2}c^{2}d_{i}}{G^{3/2}m_{i}^{2}},  \label{defparam}
\end{equation}%
where $S_{i}$ are the magnitudes of the spin vectors, $d_{i}$ are the
magnitudes of the magnetic dipole moments and $\mu _{0}=4\pi \times 10^{-7}$
Tm/A is the vacuum permeability in SI units. The magnitudes of spin vectors (%
$S_{i}=\mathcal{I}_{i}\Omega _{i}$, where $\mathcal{I}_{i}$ are the mass
moments of inertia and $\Omega _{i}$ are the spin frequencies) and the
magnitudes of magnetic dipoles $d_{i}$ appearing in these two dimensionless
parameters ($\chi _{i}$ and $\chi _{i}^{d}$) can be easily calculated for
each NS (see Table \ref{table1}). We set $a_{i}=2$ and the radius of the NS
to $10$ km in all examples below. The orbital frequencies are chosen as $20$%
, $40$, $60$ Hz, respectively, while the masses are set to $%
m_{1}=1.4M_{\odot }$ and $m_{2}=1.2M_{\odot }$, where $M_{\odot }$ is the
solar mass.

In our numerical simulation we assume the dimensionless\textit{\ }spin
parameters ($\chi _{1}=$ $\chi _{2}\equiv \chi $) and the dimensionless
magnetic dipole parameters ($\chi _{1}^{d}=\chi _{2}^{d}\equiv \chi ^{d}$)
to be equal as the companion object in a binary NS cannot always be
accurately detected in astrophysics. We set the initial individual spin and
magnetic dipole momentum vectors as $\mathbf{S_{i0}=}\chi _{i}\mathbf{(}\sin
\theta _{i}\cos \varphi _{i},\sin \theta _{i}\sin \varphi _{i},\cos \theta
_{i}$) and $\mathbf{d_{i0}=}\chi _{i}^{d}\mathbf{(}\sin \theta _{i}^{d}\cos
\varphi _{i}^{d},\sin \theta _{i}^{d}\sin \varphi _{i}^{d},\cos \theta
_{i}^{d}$) in the Cartesian coordinate system $\mathcal{K}=(\hat{x},\hat{y},%
\hat{z})$. Then, the initial spin vectors can be given in the coordinate
system with the basis of $\mathcal{K}_{S}=(\mathbf{\hat{S}_{0}^{\perp },\hat{%
J}}\times \mathbf{\hat{S}_{0}^{\perp },\hat{J}})$, where $\mathbf{%
S_{0}^{\perp }}$ is the projection of the initial total spin vector onto the
plane perpendicular to $\mathbf{J}$. The initial angles of $\theta _{i}$ ($%
\theta _{i}^{d}$) and $\varphi _{i}$ ($\varphi _{i}^{d}$) are the spherical
polar angles of the spin vector in the Cartesian coordinate system $\mathcal{%
K}$. The transformation between $\mathcal{K}_{S}$ and the Cartesian
coordinate system $\mathcal{K}$ is
\begin{equation}
A\mathcal{K}=\mathcal{K}_{S},
\end{equation}%
with the transformation matrix $A$%
\begin{equation}
A=\left(
\begin{array}{ccc}
\frac{S_{x}}{K} & \frac{S_{y}}{K} & 0 \\
-\frac{S_{y}}{K} & \frac{S_{x}}{K} & 0 \\
0 & 0 & 1%
\end{array}%
\right) ,
\end{equation}%
where $K=\sqrt{S^{2}-(\mathbf{S\cdot \hat{J})}^{2}}\ $and $S_{x}$, $S_{y}$
are the first and second components of $\mathbf{S}$. The initial spin angles
in $\mathcal{K}$ are chosen as $\theta _{1}(0)=\pi /4=\theta _{2}(0)$, $%
\varphi _{1}(0)=0$ and $\varphi _{2}(0)=\pi /2$. The initial magnetic dipole
angles in $\mathcal{K}$ are chosen as $\theta _{1}^{d}(0)=\theta _{1}(0)+\pi
/12=\theta _{2}^{d}(0)$, $\varphi _{1}^{d}(0)=\varphi _{1}(0)$ and $\varphi
_{2}^{d}(0)=\varphi _{2}(0)$. The dimensionless spin parameter is $\chi
=0.001$. We also introduce the ratio parameter $\Delta =$ $\chi ^{d}/\chi $,
which is the ratio of the dimensionless spin parameter and dimensionless
magnetic dipole parameter. Usually $\Delta \ll 1$ and is rather small for
NSs, but in the case of magnetars it can be large as $\Delta \simeq 10 $
\cite{Swift}. Thus, we investigate three cases where $\Delta =0.1,1$ and $10$
with orbital frequencies $f_{orb}=20$, $40$ and $60$ Hz. Radius $r$ in the
ASPEs can be calculated from orbital frequency using Kepler's third law as $%
r\equiv m^{1/3}(2\pi f_{orb})^{-2/3}$, where the total mass $m=m_{1}+m_{2}$.

We numerically studied the evolutions of $\cos \beta _{i}=\mathbf{\hat{J}%
\cdot \hat{S}}_{\mathbf{i}}$ for the SO, SS and QM as well as the SO, SS, QM
and DD cases. Fig. \ref{evol1} shows the evolutions of $\cos \beta _{i}$ for
$\Delta =0.1$. It can be seen that the trajectories are simple harmonic
motions, which are identical for both with and without DD effect cases. As a
result, the DD interaction can be neglected for $\Delta =0.1$. The curves in
Fig. \ref{evol1} are plotted up to $1000$ orbits and $t_{final}=1000T_{orb}$%
, where $T_{orb}=1/f_{orb}$.

Fig. \ref{evol2} shows the evolutions of $\cos \beta _{i}$ for $\Delta =1$,
while all other initial conditions are the same as in the case of Fig. \ref%
{evol1}. The dashed lines represent the DD effect. It can be seen that the
trajectories do not show simple harmonic motions from $40$ Hz upwards. The
difference between the two cases (with or without DD effect) is small with a
maximum difference in amplitude of $1\%$.

Fig. \ref{evol3} shows the evolutions of $\cos \beta _{i}$ for $\Delta =10$.
In this case the DD effect can be significant showing a $10\%$ difference in
amplitude at $20$ Hz and reaching approximately a $20\%$ difference at
frequencies over $40$ Hz.

Finally, we plotted the evolutions of angle $\cos \gamma =\mathbf{\hat{S}}_{%
\mathbf{1}}\mathbf{\cdot \hat{S}}_{\mathbf{2}}$ (Fig. \ref{evol4}). The DD
effect does not appear throughout the evolution of this angle $\cos \gamma $%
, which shows simple harmonic motion for all cases.

\section{Conclusion}

We have reviewed the leading-order orbit-averaged spin precession equations
containing the SO, SS, QM and DD interactions. This differential equation
system cannot be solved with the help of the scalar quantity $\xi $
introduced in \cite{Racine} because it is not a conserved quantity. The pure
DD case, where the SO, SS and QM contributions are neglected, can be
integrated in a simple case when the evolutions of the relative $\mathbf{%
\hat{J}\cdot \hat{S}_{i}}$ and $\mathbf{\hat{S}_{i}\cdot \hat{d}}_{\mathbf{i}%
}$ angles can be dropped. This approximation can be well applied in case of
short-term periods. In addition, we established a convenient relation
between the magnitudes of the spins and the relative angles of the magnetic
dipoles for equal mass moments of inertias.

Moreover, we have examined the full orbit-averaged spin precession equations
numerically. We demonstrated several spin configurations using numerical
investigations for some realistic binary NS systems. We introduced a
dimensionless magnetic dipole parameter to characterize the strength of
magnetic fields. Finally, we have shown that the DD effect can modify the
time evolution of the spin vectors which effect is only significant in case
of long periods of time and magnetars exhibiting large magnetic fields.

\section{ACKNOWLEDGMENT}

The work of B. M. was supported by the János Bolyai Research Scholarship of
the Hungarian Academy of Sciences.

\qquad

\end{document}